\documentclass[useAMS,usenatbib,aas_macros]{mn2e}

\usepackage{epsfig}
\usepackage{amssymb}
\usepackage{natbib}
\usepackage{aas_macros}
\usepackage{rotating}

\DeclareGraphicsExtensions{.eps}

\newcommand{\til}{~}
\newcommand{\figu}{Fig.~}
\newcommand{\figus}{Figs.~}
\newcommand{\eq}{Eq.~}
\newcommand{\sect}{Section~}

\newcommand{\tdyne}{t_{\rm dyn}}
\newcommand{\tdf}{$t_{\rm df}$}
\newcommand{\tdfe}{t_{\rm df}}

\newcommand{\forb}{$f_{\rm orb}$}

\newcommand{\forbcl}{$f_{\rm orb}/c$}

\newcommand{\bt}{$B/T$}
\newcommand{\BT}{$B/T$}

\newcommand{\gammar}{$\gamma$}

\newcommand{\mhe}{M_{\rm h}}
\newcommand{\rh}{$R_H$}
\newcommand{\re}{$R_e$}
\newcommand{\rnew}{R_{\rm new}}
\newcommand{\rbulgee}{R_{\rm bulge}}
\newcommand{\rdisce}{R_{\rm disc}}
\newcommand{\rbulge}{$R_{\rm bulge}$}
\newcommand{\rdisc}{$R_{\rm disc}$}

\newcommand{\fgase}{f_{\rm gas}}
\newcommand{\mstar}{$M_{\rm star}$}

\newcommand{\mhalo}{$M_{\rm halo}$}
\newcommand{\mhaloe}{M_{\rm halo}}
\newcommand{\mbulge}{$M_{\rm bulge}$}
\newcommand{\mdisce}{M_{\rm disc}}
\newcommand{\mdisc}{$M_{\rm disc}$}

\newcommand{\msune}{M_{\odot}}

\newcommand{\mstare}{M_{\rm star}}
\newcommand{\mbulgee}{M_{\rm bulge}}

\newcommand{\msunh}{$M_{\odot}/h$}

\begin{document}

\def\sarc{$^{\prime\prime}\!\!.$}
\def\arcsec{$^{\prime\prime}$}
\def\arcmin{$^{\prime}$}
\def\degr{$^{\circ}$}
\def\seco{$^{\rm s}\!\!.$}
\def\ls{\lower 2pt \hbox{$\;\scriptscriptstyle \buildrel<\over\sim\;$}}
\def\gs{\lower 2pt \hbox{$\;\scriptscriptstyle \buildrel>\over\sim\;$}}

\title[Sizes and Environment]{Environmental dependence of bulge-dominated galaxy sizes in hierarchical models
of galaxy formation. Comparison with the local Universe.}

\author[F. Shankar et al.]
{Francesco Shankar$^{1}$\thanks{E-mail:$\;$F.Shankar@soton.ac.uk},
Simona Mei$^{1,2}$, Marc Huertas-Company$^{1,2}$, Jorge Moreno$^{3}$,
\newauthor
Fabio Fontanot$^{4,5}$, Pierluigi Monaco$^{6,7}$, Mariangela Bernardi$^{8}$, Andrea Cattaneo$^{9}$,
\newauthor
Ravi Sheth$^{10,8}$, Rossella Licitra$^{1,2}$, Lauriane Delaye$^{1}$, Anand Raichoor$^{1}$
\\
$1$ GEPI, Observatoire de Paris, CNRS, Univ. Paris Diderot, 5 Place Jules Janssen, F-92195 Meudon, France\\
$2$ University Denis Diderot, 4 Rue Thomas Mann, 75205 Paris, France\\
$3$ Department of Physics and Astronomy, University of Victoria, Victoria, British Columbia, V8P 1A1, Canada\\
$4$ Heidelberger Institut f\"{u}r Theoretische Studien (HITS), Schloss-Wolfsbrunnenweg 35, 69118 Heidelberg, Germany\\
$5$ Institut f\"{u}r Theoretische Physik, Philosophenweg 16, 69120 Heidelberg, Germany\\
$6$ Dipartimento di Fisica - Sezione di Astronomia, Universit\`{a} di Trieste, via Tiepolo 11, I-34131 Trieste  Italy\\
$7$ I.N.A.F, Osservatorio di Trieste, Via Tiepolo 11, I- 34131, Trieste, Italy\\
$8$ Department of Physics and Astronomy, University of Pennsylvania, 209
South 33rd St, Philadelphia, PA 19104\\
$9$ Laboratoire dAstrophysique de Marseille, UMR 6110 CNRS, Univ. dAix-Marseille, 38 rue F. Joliot-
Curie, 13388 Marseille cedex 13\\
$10$ The Abdus Salam International Center for Theoretical Physics, Strada Costiera 11, 34151 Trieste, Italy}

\date{}
\pagerange{\pageref{firstpage}--
\pageref{lastpage}} \pubyear{2013}
\maketitle
\label{firstpage}

\begin{abstract}
We compare state-of-the-art semi-analytic models
of galaxy formation as well as advanced sub-halo abundance matching models
with a large sample of early-type galaxies from SDSS at $z<0.3$.
We focus our attention on the dependence of median sizes of central galaxies on host halo mass.
The data do not show any difference in the structural properties of
early-type galaxies with environment, at fixed stellar mass.
All hierarchical models considered in this work instead tend to predict a moderate to strong
environmental dependence, with the median size increasing by a factor of $\sim 1.5-3$ when moving from low
to high mass host haloes. At face value the discrepancy with the data is highly significant, especially
at the cluster scale, for haloes above $\log \mhaloe \gtrsim 14$.
The convolution with (correlated) observational errors reduces some of the tension.
Despite the observational uncertainties, the data tend to disfavour hierarchical models
characterized by a relevant contribution of disc instabilities to the formation of spheroids,
strong gas dissipation in (major) mergers,
short dynamical friction timescales, and very short quenching timescales in infalling satellites.
We also discuss a variety of additional related issues, such as the slope and
scatter in the local size-stellar mass relation, the fraction of gas in local early-type galaxies,
and the general predictions on satellite galaxies.
\end{abstract}


\section{Introduction}
\label{sec|intro}

Early-type galaxies in the local Universe are observed
to follow a rather tight size-stellar mass relation,
with an intrinsic scatter of less than a factor of two \citep[e.g.,][]{Bernardi11a,Bernardi11b,Nair11}.
This basic observational feature still represents a challenge
for hierarchical models of galaxy formation that form and evolve spheroidal systems
out of a sequence of continuous and chaotic minor and major mergers,
possibly creating scaling relations similar in slope but more dispersed
\citep[e.g.,][]{Nipoti08,ShankarPhire,Shankar13}.

On more general grounds, the different
location on the size-mass plane of galaxies at their birth \citep[e.g.,][]{ShankarBernardi09,vanderwel09,ShankarRe,Poggianti13},
as well as their environment at later times \citep[e.g.,][]{Valentinuzzi10a}, may naturally imprint
different evolutionary paths and thus different sizes to galaxies of similar
stellar mass, further contributing to enhance the expected final dispersion in scaling relations.
In hierarchical models up to 80\% of the final stellar mass
of massive bulge-dominated galaxies is predicted to be assembled via a sequence of major and minor mergers
\citep[e.g.,][]{DeLucia06,DeLucia11,Fontanot11,Khochfar11,Shankar13,Wilman13}.
Minor mergers, in particular, have been proposed as a possible driver for the size expansion
of the most massive early-type galaxies from compact, red nuggets
to the large ellipticals in the local Universe \citep{Naab09,VanDokkum10}. Possibly being more frequent in denser
environments, mergers are then believed to naturally produce larger galaxies with respect to similarly
massive counterparts in the field \citep[e.g.,][and references therein]{Shankar13}.
However, although this conjecture has been put forward in the literature
\citep[e.g.,][]{Cooper12}, it still needs to be properly verified in the context of extensive hierarchical galaxy formation models,
a task we start exploring in this work.

On the observational side, studies have recently focused
on the environmental dependence of the mass-size relation for early-type galaxies,
going from the local Universe \citep[e.g.,][]{Guo09,Weinmann09,Maltby10,Valentinuzzi10a,Huertas13a,Poggianti13},
up to $z \sim 1-2$ \citep[e.g.,][]{Valentinuzzi10b,Cooper12,Mei12,Raichoor12,Delaye13,Huertas13a,Strazzu13}.
In the local Universe several groups tend to confirm the absence
of any environmental dependence \citep[e.g.,][]{Guo09,Weinmann09,Huertas13a},
at least for massive ($\mstare \gtrsim 10^{11}\, \msune$) early type galaxies.
Some studies find cluster early-type galaxies being slightly smaller than field ones \citep[e.g.,][]{Valentinuzzi10a,Poggianti13}.
One caveat, however, is that a large fraction of lenticulars is contained
in galaxy cluster samples (see, e.g., Table 2 in \citealt{Poggianti13}).
Lenticulars tend to appear more compact at fixed stellar
mass \citep{Maltby10,Bernardi13,Huertas13a}, thus possibly influencing the analysis of samples
with significant contaminations from this latter type of galaxies.

Despite some minor observational issues which still need to be clarified,
any size increase with environment (labelled by halo mass)
seems to be overall quite negligible in the local Universe,
at least for massive early-type galaxies ($\mstare >10 ^{11}\, \msune$). This
may pose an interesting observational challenge for hierarchical galaxy evolution models,
which would naively predict a stronger galaxy growth in denser environments.

At higher redshifts, there is instead growing evidence for a possibly accelerated structural evolution of massive
early-type galaxies in very dense, cluster environments.
Preliminary studies \citep[e.g.,][]{Rettura10,Raichoor12} claimed for broadly
similar or slightly different optical morphologies for early-type galaxies in the cluster and in the field.
Using the larger and more uniform sample of galaxies extracted
from the HAWK-I cluster survey at $0.8 < z < 1.5$, \citet{Delaye13}
find instead that early-type galaxies living in clusters are about 50\% larger than
equally massive counterparts in the field (but see also \citealt{Newman13}).
\citet{Papovich12}, \citet{Bassett13}, \citet{Lani13}, and \citet{Strazzu13}
find larger galaxies with respect to the field in (proto) clusters at comparable or even higher redshifts $z \sim 1-2$,
and \citet{Cooper12} at intermediate redshifts $0.4 < z < 1.2$ in DEEP data also claimed larger early-type galaxies in denser environments.

Understanding the degree of redshift evolution of early-type galaxies
in different environments is beyond the scope of the present work.
Here we will mainly focus on model predictions and data at $z=0$, where
the statistics is much higher and at least some of the measurements more secure.
We defer the comparison to higher redshift data in separate work (Shankar et al., in prep.).
The aim of this paper is to carefully re-analyze the predictions of state-of-the-art hierarchical
semi-analytic (SAMs) and semi-empirical models of galaxy formation
with respect to their predictions on bulge sizes, and their dependence on environment (halo mass) in the local Universe.
By comparing different models developed under different techniques and physical assumptions,
the goal is to discern under which conditions the models can better line up with the data.
We note that interesting alternatives or more general interpretations that
do not necessarily rely on solely (dry) merging,
have been discussed in the literature to evolve early-type galaxy sizes
\citep[e.g.,][and references therein]{Fan10,Chiosi12,Carollo13,Ishi13,Posti13,Stringer13},
but we will reserve the investigation of these models for future studies.

The structure of the paper is as follows. We start by briefly introducing the
data sets used as a comparison in \sect\ref{sec|Data}. We then proceed
by introducing the main features of the reference models adopted in this study in \sect\ref{sec|Models}.
Our main results are then presented in \sect\ref{sec|Results}, and further discussed, along
with other caveats, in \sect\ref{sec|discu} and the Appendices. We conclude in \sect\ref{sec|conclu}.

\section{Data}
\label{sec|Data}

The early-type galaxy sample used as the reference data in this study is the one
collected and studied in \citet{Bernardi13} and \citet{Huertas13b}, and we refer
to those papers for full details on image fitting and morphological classification.
We here briefly recall that galaxies are extracted from the Sloan Digital Sky Survey
DR7 spectroscopic sample \citep{2009ApJS..182..543A},
with an early-type morphology and redshift $0.05<z<0.2$ based on the
Bayesian Automated morphological Classifier by \citet{Huertas11}.
The latter performed the automated classification of the full SDSS DR7
spectroscopic sample based on support vector machines, and associated to every galaxy
a probability to be in four morphological classes (E, S0, Sab and Scd).
Early-type galaxies are defined as those systems with a probability \verb"PROBE" to be early-type
(elliptical-E or lenticular-S0) greater than 0.5.
We note that the results are not significantly altered
if we select galaxies based only on the probability for only ellipticals (\verb"PROBELL") or
for ellipticals plus lenticulars (\verb"PROBELL"+\verb"PROBS0"). This is expected, given that central,
bulge-dominated galaxies, especially in the range of interest to this work
($\mstare \gtrsim 2\times 10^{11}\, \msune$), tend to be dominated by ellipticals.

Halo masses are taken from the group and cluster galaxy catalogue by \citet{Yang07}, updated to the DR7.
As in \citet{Huertas11}, we restricted the analysis to groups with $z < 0.09$
(for completeness reasons) and at least two members, and also removed those objects
affected by edge effects ($fedge < 0.6$). This selection ensures that $\sim 80\%$
of the groups have $\lesssim 20\%$ contamination from interlopers.
On the assumption of a one-to-one relation (with no scatter), \cite{Yang07} assigned halo masses
via abundance matching, i.e.,
via rank ordering between the total galaxy luminosity/stellar and halo mass
functions. In the specific, we use as halo mass estimate those based on the
characteristic luminosity of the group.
The expected uncertainties on such halo masses are $\sim 0.2-0.3$ dex \citep{Yang07}.

Galaxy sizes are circularized effective radii obtained from the 2D S\'{e}rsic fits
performed by \citet{Bernardi12} using the PyMorph package \citep{Vikram10},
which can fit seeing convolved two components models to observed surface brightness profiles.
Stellar masses have been obtained from the MPA-JHU DR7 release,
derived through Spectral Energy Distribution fitting using the \citet{BC03} synthesis population models,
and converted to a \citet{Chabrier03} Initial Mass Function (IMF), in line with the theoretical models described below.

Our team has actively explored the structural properties of early-type galaxies
at both low and high-redshifts. We here summarize some of our previous
empirical results relevant to the present work.
In \citet{Bernardi12} we quantified the systematics in the size-luminosity relation of galaxies in the
SDSS main sample which arise from fitting different 1- and 2-component model profiles to the images.
In particular, we emphasized that despite the half-light radius
can vary with respect to different types of fitting,
the global net effect on the R-L relation
is small, except for the most luminous tail, where it curves upwards towards larger sizes.
Compared to lower mass galaxies and previous work in the local Universe, the slope
is in fact $\beta \sim 0.85$ instead of the commonly reported slope of $\beta \sim 0.5-0.6$
\citep[e.g.,][]{Shen03,Cimatti12}. This difference is mainly due to the way
\citet{Bernardi12} fit the light profile, and in part to the sky subtraction.
We will further expand on this point in \sect\ref{subsec|SizeMstarRelation}.
In \citet{Huertas13a} we used the above defined sample of $z \sim 0$ SDSS early-type galaxies
to point out a negligible dependence of the sizes on environment, at fixed stellar mass.
More specifically, we were able to demonstrate via detailed
Monte Carlo simulations that considering our observational errors and the size of the sample,
any size ratio larger than $30-40\%$ between massive galaxies ($\log \mstare/\msune>11$)
living in clusters and in the field could be ruled out at $3\sigma$ level.
The analysis yielded similar results irrespective of the explicit galaxy selection,
either on type (central/satellite), star formation rate, exact early-type morphology, or central density,
at least for galaxies above $\gtrsim 10^{11}\, \msune$.
In the same work we also emphasized that our findings
on a null dependence on environment were not induced by a galaxy sample biased
towards possibly more evolved systems with higher values
\footnote{In hierarchical scenarios, for example, more evolved systems, i.e., with more mergers, could be expected
to have, on average, higher values of the the S\'{e}rsic index \citep[e.g.,][]{Hop08FP}.} of the S\'{e}rsic index $n$.
Our early-type galaxy sample is in fact characterized by broad S\'{e}rsic index distributions, with
a slight dependence on stellar mass.
More quantitatively, one could broadly define a linear relation
of the type $n$-$\log \mstare$, with a slope of $\sim 0.8$ and scatter of $\sim 1.2$.
We will further discuss the negligible environmental dependence of the size-stellar mass
relation in SDSS early-type galaxies in \sect\ref{subsec|comparedata}.
In the following, we will use this large and accurate galaxy sample of early-type galaxies
as a base to compare with detailed predictions from a suite of semi-analytic and semi-empirical models
presented in the next Section.

\section{Theoretical Models}
\label{sec|Models}

Before entering into the details of each galaxy formation model adopted in this work,
we first summarize some key, common properties of how the mass and structure of bulges
are evolved in hierarchical models.
Clearly, models include a variety of physical processes, including gas cooling,
supernova feedback, stellar/gas stripping, super-massive black hole feeding and feedback, etc..
and we defer the reader to the original model papers (cited below) for complete details on their full implementations.
In the following, we will mainly focus on those physical processes which have a direct impact on shaping bulge sizes.

Galaxies evolve along dark matter merger trees via in-situ star formation
and mergers from incoming satellites.
Galaxies are usually assumed to initially have a disc morphology via conservation of specific angular momentum,
and then evolve their morphology via mergers and disc instabilities.
When galaxies become satellites in larger haloes, they are assigned
a dynamical friction timescale \tdf\ for final coalescence with the central galaxy
\begin{equation}
\tdfe=\tdyne T(\mhaloe/M_s,\rm{orbit})\, ,
    \label{eq|tdynGeneral}
\end{equation}
where $T$ is a general function of the mass ratio between main dark matter halo
\mhalo\ and the satellite $M_s$, as well as the orbital parameters.
The dynamical timescale is defined as $\tdyne=0.1 H(z)^{-1}$, where $H(z)$ is the Hubble's parameter.
Each model generally adopts a somewhat different analytic treatment
for \tdf, which in turn has an impact on the cumulative rate of mergers
per galaxy. In the following, we will only briefly emphasize the key differences relevant to our discussion.
Full details on the comparison of dynamical friction timescales among different models
can be found in, e.g., \citet{DeLucia10}.

When a merger between a central and a satellite galaxy actually occurs,
models broadly distinguish two possibilities.
In violent {\em major} mergers (in which the ratio of the baryonic masses of the progenitors is usually
assumed to be $M_2/M_1>0.3$), discs are completely destroyed forming a spheroid\footnote{None of the models considered in this work
include disc survival after a major merger, even if the
merger is sufficiently gas-rich. However,
this is believed to be an important aspect
only when dealing with the evolution of more disc dominated,
less massive systems, such as lenticulars.
Disc survival is believed to play a relatively minor role for the bulge dominated massive ($\gtrsim 2\times 10^{11}\, \msune$)
galaxies of interest here, with relatively minor
gas leftover after the major merger, and late mass assembly dominated by minor, dry mergers.
For the latter systems, the half-mass radius is largely dominated by the bulge component, as
also empirically confirmed from detailed bulge-to-disc decompositions morphological fitting \citep{Bernardi13}.
Recent semi-analytic modelling \citep{DeLucia11,Wilman13} confirm disc survival
to be a non-negligible component mainly for low to intermediate masses, and at high redshifts. We will
anyway discuss disc survival where relevant.}.
The remnant's stellar mass is then composed of the
stellar mass of the progenitors and a given fraction, depending on the model,
of the gas present in the merging discs,
properly converted into stars in a burst.
In {\em minor} mergers ($M_2/M_1<0.3$), the stars
of the accreted satellite are added to the bulge of the central
galaxy, while any accreted gas can be either added to the main gas disc,
without changing its specific angular momentum, or converted to stars
and added to the bulge, according to the model, as detailed below.

Particularly relevant for the present study is the computation of bulge sizes.
We summarize in Table\til\ref{table|parameters} all the key physical parameters
adopted in the hierarchical models considered in this work, playing a significant
role in shaping the size distribution of bulges and spheroids.
A description of the relevant processes and related parameters is given below.
For the rest of the paper we will mainly focus our attention on bulge-dominated
galaxies with \bt$>0.5$, although we will discuss the effects
of tighter cuts in the selection where relevant.

\citet{Cole00} were the first to include in their model an analytic treatment
of bulge sizes, and all the other hierarchical galaxy formation models
considered here followed their initial proposal. The size of the remnant $\rnew$
is computed from the energy conservation between the sum of the self-binding energies of the
progenitor galaxies, and that of the remnant \citep{Cole00}
\begin{equation}
\frac{(M_1+M_2)^2}{\rnew}=\frac{M_1^2}{R_1}+\frac{M_2^2}{R_2}+\frac{f_{\rm orb}}{c}\frac{M_1M_2}{R_1+R_2}\, ,
\label{eq|sizegeneral}
\end{equation}
where $M_i$, $R_i$, are, respectively, the total masses and half-mass radii of the merging
galaxies. The form factor $c$, depends weakly on the galaxy density profile varying
from 0.45 for pure spheroids to 0.49 for exponential discs \citep{Cole00}.
The factor $f_{\rm orb}$ instead parameterizes the (average) orbital energy of the merging systems,
ranging from zero for parabolic orbits, to unity
in the limit in which the two pre-merging galaxies are treated as point masses in a circular orbit with separation $R_1+R_2$.
Effectively, the ratio \forbcl\ can be considered as a free parameter.

\eq\ref{eq|sizegeneral} does not include gas dissipation which, as revealed
by high-resolution hydro-simulations \citep[e.g.,][and references therein]{Hop08FP,Covington11},
tends to shrink bulges formed out of gas-rich mergers, more than
what would be predicted by the dissipation-less mergers defined
in \eq\ref{eq|sizegeneral}. \citet{Hop08FP} proposed a rather
simple prescription to include gas dissipation in mergers as
\begin{equation}
\rnew=\frac{\rnew[dissipationless]}{1+F_{\rm gas}/f_0}\, ,
\label{eq|Redissipation}
\end{equation}
where $f_0=0.25-0.30$, $F_{\rm gas}$ is the ratio between the total mass of cold gas
and the total cold plus stellar mass (inclusive of the mass formed during the burst) of the progenitors,
and $\rnew[dissipationless]$ is computed from \eq\ref{eq|sizegeneral}.
We will discuss the impact of gas dissipation in the relation between size and environment.

In most of the models bulges are also assumed to grow via disc instabilities.
The general criterion adopted for disc instability in the SAMs discussed here
is expressed as \citep{Efstathiou82}
\begin{equation}
\epsilon > \frac{V_{\rm ref}}{\sqrt{GM_{\rm disc}/R_{\rm disc}}}\, ,
\label{eq|discinstability}
\end{equation}
with the circular velocity of the disc expressed in terms of its mass $M_{\rm disc}$ and
half-mass radius $R_{\rm disc}$ (for exponential profile, equal to 1.68$R_D$, with $R_D$ the disc scale-length).
The reference velocity $V_{\rm ref}$ is usually expressed as a linear function of the
circular velocity of the host halo or the disc itself, while $\epsilon$ is a real number of order unity, as detailed below.
When the circular velocity of the disc becomes larger than
a given reference circular velocity, then the disc is considered unstable
and mass is transferred from the disc to the bulge.
\eq\ref{eq|discinstability} expresses the physical condition that when the disc becomes sufficiently
massive that its self-gravity is dominant, then it tends to be unstable to any small
perturbation.

In the case of disc instabilities the size of the bulge is also computed via an energy conservation
equation \citep{Cole00} equivalent to \eq\ref{eq|sizegeneral}
\begin{equation}
\frac{(\mbulgee+\mdisce)^2}{\rnew}=\frac{\mbulgee^2}{\rbulgee}+\frac{c_D}{c_B}\frac{\mdisce^2}{\rdisce}
+\frac{f_{\rm int}}{c_B}\frac{\mbulgee\mdisce}{\rbulgee+\rdisce}
\label{eq|sizediscinstab}
\end{equation}
which expresses a merger-type condition between the unstable disc
with mass \mdisc\ and half-mass radius \rdisc, and any pre-existing bulge
with mass \mbulge\ and half-mass radius \rbulge.
Following \citet{Cole00}, all models below use the values of $c_B\sim c_D \sim 0.5$, for the bulge and disc form factors,
and $f_{\rm int}=2$ for the constant parameterizing the gravitational interaction term between the disc and the bulge.
As discussed by \citet{Guo11}, a higher value of $f_{\rm int}=2$ for the interaction term with respect to
the value of $f_{\rm orb}\lesssim 1$ usually used in \eq\ref{eq|sizegeneral},
physically takes into account that the interaction in concentric shells is stronger than in a merger.
This in turn implies that for similar stellar mass of the remnant bulge,
a disc instability will inevitably produce more compact sizes with respect to a merger.
In other words, in this formalism
mergers are considered to be more efficient in building larger bulges and spheroids.

\begin{table*}
\begin{tabular}{|l|c|r|}
  \hline
  \hline
  \emph{PARAMETER} & \emph{DESCRIPTION} & \emph{VALUE} \\
  \hline
  \hline
  $\mu=M_{\rm sat}/M_{\rm cen}$ & Ratio between the baryonic masses of & 0.3 \\
  & satellite and central. Mass ratios above/below this  & \\
  &  threshold are treated as major/minor merging  & \\
  \hline
  $e_{\rm burst}$ & Fraction of cold gas converted to stars & 0-1 \\
                  & in a merging and added to the bulge  &  \\
  \hline
  $(\tdfe/\tdyne)/(\tdfe/\tdyne)_{\rm num\, sims}$ & Ratio of the dynamical friction timescale in units & 0.1-1 \\
  & of the dynamical time adopted in models,
  compared to  & \\
  & that from controlled numerical simulations & \\
  & (see Fig. 14 in \citealt{DeLucia10}) & \\
  \hline
  $f_{\rm orb}$ & Average orbital energy of the merging systems  & 0-1 \\
  \hline
  $\rnew[dissipation]/\rnew[dissipationless]$ & Ratio between size of the remnant & $\sim 0.1-1$ \\
   & in the dissipation and dissipationless case &  \\
  \hline
  $f_{\rm int}$ & Gravitational interaction term & 2 \\
  & between the disc and the bulge & \\
  \hline
  $\epsilon$ & Ratio between reference circular velocity& $\sim 1$ \\
   & and the circular velocity of the disc &  \\
  \hline
  \hline
  \end{tabular}
  \caption{List of the main parameters adopted in the hierarchical galaxy formation models discussed in this work
  responsible for shaping the sizes of bulges and spheroids.}
  \label{table|parameters}
\end{table*}

\subsection{The Durham model by Bower et al. (2006)}
\label{subsec|B06model}

One popular rendition of the Durham galaxy formation
models\footnote{Available at http://www.g-vo.org/MyMillennium3.} is the one by \citet[][B06 hereafter]{Bower06}.
This model is built on the Millennium I simulation \citep{Springel05},
composed of $N= 2160^3$ dark matter particles of mass $8.6\times10^{8}\,h^{-1}{\rm M}_{\odot}$,
within a comoving box of
size $500\, h^{-1}$Mpc on a side, from $z=127$ to the present, with cosmological parameters
$\Omega_{\rm m}=0.25$, $\Omega_{\rm b}=0.045$, $h=0.73$,
$\Omega_\Lambda=0.75$, $n=1$, and $\sigma_8=0.9$.

Galaxies in this models are self-consistently evolved
within merger trees which differ with respect to the original ones
presented by \citet{Springel05},
both in the criteria for identifying independent haloes,
and in the treatment and identification of the descendant haloes
\citep[see details in][]{Harker06}.
The dynamical friction timescales
adopted by B06 follow \citet{Cole00} and, as shown in \citet{DeLucia10}, they can be factors of
$\gtrsim 2-3$ to $\gtrsim 10$, respectively for major and minor mergers, lower than
those extracted from controlled numerical, high-resolution cosmological simulations
\citep[e.g.,][]{Boylan08}.

In a major merger, following \citet{Cole00}, B06 assume a single bulge or elliptical
galaxy is produced, and any gas present in the discs of the merging
galaxies is converted into stars in a burst.
In a minor merger, all the stars of the accreted satellite are added to the bulge of the central
galaxy, while the gas is added to the main gas disc.
In \eq\ref{eq|discinstability} B06 set $V_{\rm ref}$ as the
circular velocity at the half-mass radius of the disc, with
$\epsilon \sim 1$, and assume that when the disc goes unstable the entire mass of the disc is
transferred to the galaxy bulge, with any gas present assumed to undergo a starburst,
and adopt the values of $c_B=0.45$, $c_D=0.49$ for the bulge and disc form factors.

Finally, following \citet{Cole00},
B06 also include some halo adiabatic contraction prescriptions that slightly
modify the sizes as calculated out of \eq\ref{eq|sizegeneral} and \eq\ref{eq|sizediscinstab},
but the effects of these re-adjustments are relatively small \citep[e.g.,][]{Gonzalez09}.

\subsection{The Munich model by Guo et al. (2011)}
\label{subsec|G11model}

One of the latest renditions of the Munich model\footnote{Available at http://www.g-vo.org/MyMillennium3.} has been
published in \citet[][G11 hereafter]{Guo11} ,
and we use their run on the Millennium I simulation (with merger trees
from \citealt{Springel05}).
The satellite total infall time is given by
the destruction time of the subhalo due to tidal truncation and stripping,
plus an additional dynamical friction timescale
down to the coalescence of the subhalo with the centre of the main
halo. Overall, the Munich total merging timescales are comparable to, although
in extreme minor merging regime somewhat shorter than,
those from high-resolution cosmological simulations \citep{DeLucia10}.

The G11 model evolves gas and stellar discs in an inside-out
fashion, adding material to the outskirts following conservation of angular momentum.
\citet{Guo11} have shown that their model is capable of reproducing
the size distribution of local discs reasonably well \citep[additional
comparisons can be found in, e.g.,][]{Fu10,Kauff12,Fu13}.

As in B06, G11 assume that in minor mergers the pre-existing
stars and the gas of the satellite are added to
the bulge and to the disc of the primary galaxy, respectively.
G11 also allow for some new stars to be formed during
any merger following the collisional starburst model by \citet{Somerville01},
where only a fraction
\begin{equation}
e_{\rm burst} = 0.56 \left(\frac{M_{\rm 2}}{M_{\rm 1}}\right)^{0.7}
\label{eq|eburst}
\end{equation}
of the cold gas of the merging galaxies is converted into stars.
The new stars are then added to the bulge or to the disc, depending on the merger
begin major or minor, respectively.

When computing bulge sizes, the G11 model also takes into account the fact that only
the stellar bulge of the central partakes in a minor merger with the satellite, thus
$M_1$ and $R_1$ in \eq\ref{eq|sizegeneral} are replaced by the bulge mass and half-mass radius,
respectively. In a major merger,
G11 limit the virial masses $M_1$ and $M_2$ entering \eq\ref{eq|sizegeneral}
to the sum of stellar mass plus the fraction of gas converted into stars, assumed
to be distributed with an exponential profile with half-mass
radius computed following the full prescriptions given in G11.
G11 also adopt a fiducial value of \forb$=0.5$ in \eq\ref{eq|sizegeneral}.

The disc instabilities are treated somewhat differently in the G11 model.
First, in the condition for instability in \eq\ref{eq|discinstability}, G11
set $\epsilon=1/\sqrt3$ and $V_{\rm ref}$ equal to the maximum circular velocity of the (sub)halo.
Second, when a disc goes unstable, only the necessary fraction of stellar mass $\delta \mstare$ in the
disc is transferred to the bulge to keep the system marginally stable.
Third, G11 adopt \eq\ref{eq|sizediscinstab} to compute bulge sizes in disc instabilities
\emph{only} if a bulge is already present. If not, then it is assumed that the mass
$\delta \mstare$ is transferred, with no loss of angular momentum,
from the inner part of the disc (with the exponential-like
density profile) to the forming bulge, in a way that the bulge half-mass radius equals
the radius of the destabilized region
\begin{equation}
\delta \mstare=2\pi \Sigma_0 R_D [R_D-(R_D+\rbulgee)\exp(-\rbulgee/R_D)]\,
\label{eq|initialradiusdiscGuo}
\end{equation}
where \rbulge\ is the half-mass radius of the newly formed bulge, and $\Sigma_0$
the central density of the disc.

\subsubsection{Modifications to the Guo et al. (2011) model}
\label{subsec|S13model}

The G11 model does not include gas dissipation.
\citet{Shankar13} have modified the G11 numerical
code to include gas dissipation during \emph{major} mergers as
given in \eq\ref{eq|Redissipation}.
They also adopted \forb$=0$ together with dissipation, as this combination yielded an
improved match to the local size-stellar mass relation.
We will discuss the impact of this variant
of the G11 model to the general predictions on environment, and label this model as S13 in the following.

\subsection{The {\sc morgana} model}
\label{subsec|MORGANA}

The {\sc morgana} model uses as an input the dark matter
merger trees obtained with the PINOCCHIO algorithm \citep{Monaco02}.
This does not give information on halo substructures. In the original version
of {\sc morgana} galaxy merging times are computed
using the model of \citet{Taffoni03},
which takes into account dynamical friction, mass loss by tidal stripping,
tidal disruption of substructures, and tidal shocks. However, the \citet{Taffoni03}
timescales have been shown to be significantly shorter than
those obtained from N-body simulation by, e.g., \citet{Boylan08}.

In this work we will use a version of {\sc morgana}
presented in \citet{DeLucia11} and \citet{Fontanot11}.
This implements longer dynamical friction timescales for satellites,
consistent with those of \citet{DeLuciaBlaizot}.
In addition, this version of the model
does not include the scattering of stars to the diffuse stellar component of
the host halo that takes place at
galaxy merging \citep{Monaco06}.
This is particularly relevant for this paper as it maximizes the
effect of mass growth via mergers, because satellites retain all their mass
before final coalescence thus allowing a more efficient size growth
in the remnant (cfr. \eq\ref{eq|sizegeneral}).

Mergers and disc instabilities move mass from the disc to the bulge component
through very similar analytic prescriptions as the ones in B06.
In minor mergers ($M_2/M_1<0.3$), the whole satellite is added to the bulge,
while the disc remains unaffected.
The latter characteristic boosts
the growth in mass of the centrals, rendering minor mergers
more efficient in size growth than for, e.g., the G11 model.
In major mergers, all the gas and stars of the two merging
galaxies are given to the bulge of the central one.
Sizes in mergers follow energy conservation given in \eq\ref{eq|sizegeneral},
with \forbcl$=2$.
In addition to these processes, cooling and infall from the halo onto a bulge/disc system
is assumed to deposit cold gas in the bulge as well, for a fraction equal to the disc
surface covered by the bulge. This is done to let feedback from the central black hole
respond quickly to
cooling without waiting for a merging or disc instability. This process is
responsible for a minor part of mass growth of bulges.

For disc instabilities, {\sc morgana} uses a threshold
given by \eq\ref{eq|discinstability} with $\epsilon=0.9$, $V_{\rm ref}$
being the disc rotation velocity (computed with a model like \citealt{MMW}
which takes into account the presence of the bulge) at 3.2 scale radii.
In disc instability events this model assumes
that 50\% of the disc mass is transferred to the bulge, and the size of the forming
bulge is given by \eq\ref{eq|sizediscinstab} with $C_B=C_D=0.5$.
In this respect, the {\sc morgana} model can be considered to be midway between
the G11 model characterized by relatively weak disc instabilities, and the B06
model with maximal instabilities. To better isolate the impact
of disc instabilities on model results, in the following we will also discuss a realization
of the {\sc morgana} model with the same identical prescriptions as the one
just described but with no disc instabilities.

\subsection{Sub-Halo Abundance Matching Model (SHAM)}
\label{subsec|SHAM}

We also include in our analysis the results of
a sub-halo abundance matching model (SHAM). This approach relies on
progressively more popular semi-empirical techniques adopted to study a variety
of galaxy properties, from colours to structure \citep[e.g.,][]{Vale04,Shankar06,Hop08FP,HopkinsMergers,Leauthaud10,Bernardi11a,Bernardi11b,Neistein11,Moster13,Yang12,Watson12}.
Probing galaxy evolution via a semi-empirical model
like the one sketched in this section, allows to restrict the analysis to just a few basic
input parameters, i.e., just the ones defining the underlying chosen physical assumptions (e.g., mergers and/or disc
instabilities), as all other galaxy properties are fixed from observations.

Our model starts from 20,000 dark matter merger trees
randomly extracted from the Millennium simulation, but uniformly\footnote{When computing statistical
distributions of any quantity extracted from the SHAM
model, we will always include proper galaxy weights. The latter
are given by computing the ratio between the integral of the halo
mass function over the volume of the Millennium Simulation
and over the bin of halo mass considered, divided by the number
of galaxy hosts in the Monte Carlo catalog in the same bin.
We note that even ignoring the weighting would have a negligible
impact on any result on the size distributions at fixed stellar mass.},
in the range $10^{11}$ \msunh\ to $10^{15}$ \msunh.
Inspired by the methodologies adopted by \citet{Hop08FP} and \citet{Zavala12},
a (central) galaxy inside the main progenitor branch of a tree is at each timestep initialized
in all its basic properties (stellar mass, gas fraction, structure, etc...) via empirical relations until a merger occurs.
Central galaxies are assumed to be initially gas-rich discs,
and then evolve into a spheroid via a major merger,
and/or grow an inner bulge via minor mergers and/or, possibly, disc instabilities.
After a major merger occurs, the central galaxy is no longer re-initialized
and it remains frozen in all its baryonic components, although we still allow for stellar and gas mass growth via mergers.

SHAM models have the virtue that they do not require full ab initio physical
recipes to grow galaxies in dark matter haloes, as in extensive galaxy formation models (SAMs).
This in turn allows to bypass the still substantial unknowns
in galaxy evolution about, e.g., star formation, cooling, and feedback
which in turn may drive more sophisticated galaxy formation models
to serious mismatches with basic observables such as the stellar mass function \citep[e.g.][]{Henriques12,Guo13}.
SHAM models instead use the stellar mass function and other direct observables
as \emph{inputs} of the models,
allowing us to concentrate on other galaxy properties,
such as mergers and the role of environment, making them
ideal, complementary tools for studies such as the one undertaken here.

More specifically, we assume that initially central galaxies
are discs with an exponential profile following at all times (we here consider
the evolution at $z \le 3$, where the data are best calibrated)
the redshift-dependent \mstar-\mhalo\ relation defined by \citet{Moster13} (for a Chabrier IMF) as
\begin{equation}
\mstare=2\mhaloe N \left[\left(\frac{\mhaloe}{M_1}\right)^{-\beta}+\left(\frac{\mhaloe}{M_1}\right)^{\gamma}\right]^{-1}
    \label{eq|MstarMhaloMoster}
\end{equation}
with all the parameters $N$, $M_1$, $\beta$, and $\gamma$ varying with redshift
as detailed in \citet{Moster13}.
Despite \eq\ref{eq|MstarMhaloMoster} being an improvement with respect
to previous attempts, as it takes into account measurement
errors on the stellar mass functions, the exact correlation between stellar mass and halo
mass is still a matter of debate \citep[e.g.,][]{Neistein11,Yang12}. Nevertheless, using
other types of mappings \citep{Yang12} does not qualitatively affect our global discussion
which is mainly based on comparisons at \emph{fixed} bin of stellar mass.

Gas fractions are assigned to each central disc galaxy according to its current stellar mass
and redshift using the empirical fits by \citet{Stewart09}
\begin{equation}
\fgase=\left(\frac{\mstare}{4.5\times 10^{11}\msune}\right)^{a(z)}
    \label{eq|fgasStewart}
\end{equation}
with $a(z)=-0.59(1+z)^{0.45}$.
Disc half-mass, or half-light, radii (which we here assume equivalent, i.e., light traces mass)
are taken from the analytic fit by \citet{Shen03}
\begin{equation}
R_{\rm disc}=\frac{R_0}{(1+z)^{0.4}}\mstare^{k}\left(1+\frac{\mstare}{3.98\times 10^{10}\, \msune}\right)^{p-k}
    \label{eq|rdisk}
\end{equation}
with $R_0$=0.1, $k=0.14$, $p=0.39$
(input stellar masses in \eq\ref{eq|rdisk}, defined for a Chabrier IMF, are corrected following \citealt{Bernardi10} by 0.05 dex to
match the IMF used by Shen et al.).
The extra redshift dependence of $(1+z)^{-0.4}$ in \eq\ref{eq|rdisk} at fixed stellar mass
is adapted from, e.g.,
\citet{Somerville01} and \citet{Hop08FP}.
Although observations may provide slightly different normalizations and/or slope for \eq\ref{eq|rdisk}
(see, e.g., discussion in \citealt{Bernardi12}), this does not alter our conclusions.

After a merger we assume the mass assembly and structural growth criteria as in G11.
In a major merger the central galaxy is converted into an elliptical, with its
stellar mass equal to the sum of those of the merging progenitors as well as
the gas converted into stars during the starburst following \eq\ref{eq|eburst}.
In a minor merger only the stars of the satellite are accreted to the bulge.
Bulge sizes are determined from \eq\ref{eq|sizegeneral}.

To each infalling satellite, we assign all the properties of a central galaxy living in a typical halo,
randomly extracted from the overall Monte Carlo catalog of central galaxies, with the
same (unstripped) mass as the satellite host dark matter halo at infalling time.
Observational uncertainties in calibrating the exact morphologies of especially lower mass galaxies \citep[e.g.,][]{BakosTrujillo}
anyway still limit our true knowledge of merging progenitors,
and recent studies seem to show that the vast majority of the high redshift massive galaxies
are disc-dominated \citep[e.g.,][and references therein]{Huertas13a}.
By simply approximating all infalling satellites as discs (i.e.,
with negligible bulges), in line with what assumed by \citet{Zavala12},
any dependence of size with host halo mass would be less strong
than the ones actually presented below.

Dynamical friction timescales are taken from the recent work of \citet{McCavana12}
\begin{equation}
\tdfe=\tdyne \frac{A(\mhaloe/M_s)^{B}}{\ln (1+\mhaloe/M_s)}\exp \left[C \frac{J}{J_c(E)}\right]\left[\frac{r_c(E)}{R_{\rm vir}}\right]^D
    \label{eq|tdynMcCavana}
\end{equation}
with $A=0.9$, $B=1.0$, $C=0.6$, and $D=0.1$, but we checked that using
the values of these parameters inferred by \citet{Boylan08} instead yields similar results.
Following \citet{KhochfarBurkert06}, to each infalling satellite we assign a circularity $\eta=J/J_c(E)$
randomly extracted from a Gaussian with average 0.50 and dispersion of 0.23 dex, from which
we compute $r_c=R_{\rm vir}\eta^{2.17}/(1-\epsilon)$, with $\epsilon=\sqrt{1-\eta^2}$.

What is also relevant to size evolution of central galaxies, as further detailed below,
is how we treat satellite evolution in stellar mass and size once they fall
in more massive haloes, i.e., the degree
of (gas and star) stripping and/or the amount of residual star formation (which self-consistently grows
stellar mass and disc radius).
In our basic model, we assume in line with many observational and/or semi-empirical results
\citep[e.g.,][]{Muzzin12,Krause13,Mendel13,Woo13,Wetzel13,Yang13}, that satellite galaxies continue
forming stars according to their specific star formation rate for up to a few Gyrs.
The latter is in agreement with the recent results of \citet{Mok13}, who find
any delay between accretion and the onset of truncation of star formation to be $\lesssim 2$ Gyr,
at least for massive satellites in the range $10^{10}-10^{11} \msune$,
the ones of interest to the present work.

Satellites continue forming stars according the their availability of residual gas,
and at the rate specified by their specific star formation
at infall as \citep{Karim12,Peeples13}
\begin{equation}
{\rm SSFR}=\frac{0.0324}{\mstare}(1+z)^{3.45}\left(\frac{\mstare}{10^{11}\, \msune}\right)^{0.65} \,\,\,\,\, Gyr^{-1}\, .
    \label{eq|SFR_Karim}
\end{equation}

Note that our simple star formation prescription for satellites does not take into account
any detailed treatment of stellar feedback during the life of the satellite, but simply prolongs
in time the physical conditions at infall. In other words, we safely assume that the specific
star formation rate associated to the galaxy is the ``equilibrium one'', result of the balance between
gas infall and feedback.

Finally, for completeness, we include in the scaling relations initializing centrals and infalling satellites,
a lognormal scatter of 0.15 dex around the median stellar mass \citep{Moster13}, a mean 0.2 dex in the gas
fraction \citep{Stewart09}, a 0.1 dex in in the input $f_{\rm orb}$ parameter \citep{KhochfarBurkert06},
a 0.1 dex in specific star formation rate \citep{Karim11}, and a median 0.1 dex in disc radius \citep{Somerville08}.

To summarize, a SHAM model empirically initializes central galaxies as stellar, gas-rich discs.
Satellites are assigned all the properties of a central galaxy living in a typical halo of the
same mass of the host at the epoch of infall.
Satellite galaxies can then be quenched, and/or stripped, and/or continue to form stars
according to their SSFR. Centrals instead at all epochs continue to be updated
along the main progenitor halo in the dark matter merger trees until a merger occurs.

\subsubsection{Variants to the reference SHAM}
\label{subsubsec|SHAMmodels}

As discussed above, galaxy evolution via semi-empirical models is restricted to fewer basic
input parameters. This in turn
allows a more direct and transparent understanding
of the impact of any additional input physical process.
In the following when comparing with the data,
we will thus also discuss
several variations to our reference SHAM, alongside with the more extensive semi-analytic
models discussed above.

More specifically, we will present the following set of key
variants to the reference SHAM.
\begin{itemize}
  \item A SHAM characterized by \forb$=0$ (keeping a dispersion of 0.3 dex), i.e., assuming on average parabolic orbits.
  \item A SHAM with \forb$=0$, with gas dissipation in major mergers following \eq\ref{eq|Redissipation}.
  \item A SHAM with \forb$=0$, and satellites undergoing
fast quenching after infall (i.e., 0.5 Gyrs instead of the 2 Gyrs of the reference model).
  \item A SHAM with \forb$=0$, which adopts
a dynamical friction timescales a factor of $1/3$ less than the one by \citet{McCavana12}, used as a reference in all other SHAM models.
  \item A SHAM equal to the reference one, also including
  an empirically motivated mass-dependent stellar and gas stripping,
  parameterized as \citep{Cattaneo11}
  \begin{equation}
  {\rm F_{\rm strip}}=(1-\eta)^{\tau} \, ,
  \label{eq|stripping}
  \end{equation}
  with $\tau=\tdfe/\tdyne$ the ratio between the dynamical friction and dynamical timescales.
  As detailed below, the exact consumption of gas via star formation during infall
  is nearly fully degenerate with the amount of stripping assumed in the models.
  We will discuss the value adopted for the $\eta$ parameter in the next sections.
  Stellar stripping does not only affect stellar mass but also disc size. We assume that, on average,
  the disc during its evolution always strictly follows an exponential profile, with its central
  density obeying the relation $\Sigma=\mdisce/ 2 \pi Rs$. Thus we assume the central density to be
  conserved at each stripping event and update stellar and disc radius accordingly.
\end{itemize}

Finally, although we include in all SHAMs mild bar instabilities
following \eq\ref{eq|initialradiusdiscGuo}, we find
in our semi-empirical models
the latter process to a very minor role in the build-up of massive bulges.
We can thus safely refer to our SHAMs as models with negligible disc instabilities.
Given that the full range of disc instabilities from moderate, to strong, to very strong ones,
have already been extensively covered by the reference semi-analytic models
discussed above (G11/S13, {\sc morgana}, and B06, respectively),
we will not further pursue this issue in SHAMs.

\section{Results}
\label{sec|Results}

\subsection{Comparison Strategy: Sample selection and treatment of observational errors}
\label{subsec|comparisonstrategy}

In our comparison between galaxy models and data we will mainly focus on \emph{central} galaxies.
Central galaxies are the ones believed to be the most affected by mergers, especially at later times, and should thus be those
type of systems for which any environmental dependence is in principle maximized.
We will anyhow briefly discuss satellites in Appendix\til\ref{App|Satellites}.

We stress that in this work we preferentially select galaxies based on their morphology.
We are in fact here mainly interested in studying the global structure of galaxies as a function
of stellar mass and environment, and thus do not attempt to impose any further cut in, e.g.,
star formation rate, to limit the selection to passive galaxies. Nevertheless, we note that
the most massive local massive and central galaxies of interest here are mostly passive.
We have checked, for example, that the distributions predicted by the G11 and S13 models
are nearly unchanged if we restrict to very massive galaxies with a specific star formation
rate below, e.g., $<0.01\, {\rm Gyr^{-1}}$.
This is in line with the observational evidence reviewed in
\sect\ref{sec|Data} which suggests
the null environmental dependence to be independent of the
exact selection adopted.

\begin{figure*}
    \includegraphics[width=15truecm]{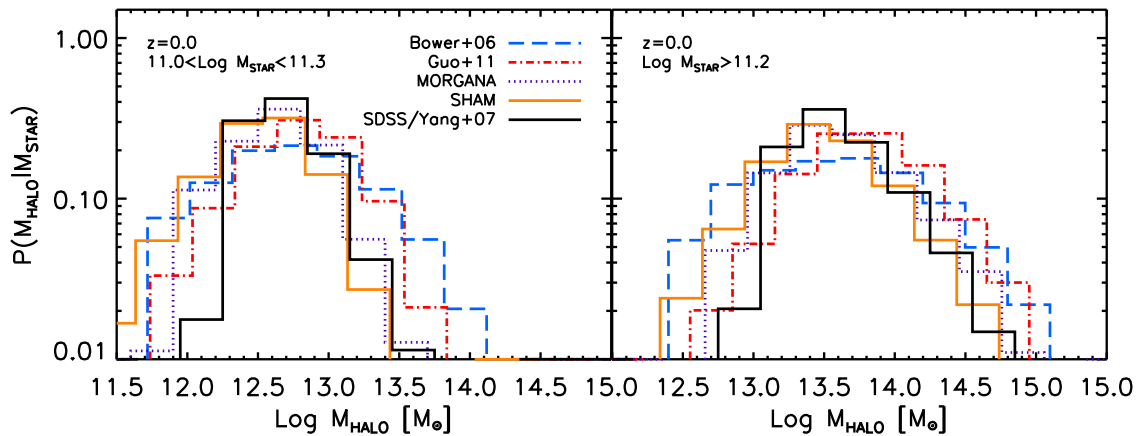}
    \caption{Predicted halo mass distributions for central galaxies of stellar mass
    in the range $11.<\log \mstare<11.3$ (\emph{left}) and $\log \mstare>11.2$ (\emph{right}), for different galaxy
    formation models, as labelled. The \emph{solid}, \emph{black} lines are the corresponding distributions for central early-type galaxies
    in SDSS (see text for details).
    All predicted and observed distributions are broad, covering about two orders of magnitude in halo mass, if not more.}
    \label{fig|PMhaloFixedMstar}
\end{figure*}

When discussing environmental trends (\sect\ref{subsec|trend}),
we chose to consider in our analysis all early-type galaxies
more massive than $\log M_1/\msune>11.2$, an interval of stellar mass
which includes galaxies up to a few $\log \mstare/\msune \sim 12$.
This solution allows the bin to be sufficiently
large not to be dominated by errors in stellar mass estimates \citep[e.g.,][]{Huertas13a},
and at the same time to maximize the statistics in about three decades of host halo mass (cfr. right panel
of \figu\ref{fig|PMhaloFixedMstar}), from the field/small group scale with $\log \mhaloe/\msune \gtrsim 12.5$,
to massive clusters with $\log \mhaloe/\msune \lesssim 15$. We stress, however, that varying the
amplitude of the stellar mass interval does not qualitatively impact the general trends discussed in
this work \citep[see, e.g.,][]{Huertas13a,Huertas13b}.

When comparing model predictions to data we will start off by showing the raw model predictions.
However, we are here interested to probe not only
the median correlations, but even more to understand the scatters
around these relations, and how the latter depend on environment.
It is thus essential to take into account observational errors,
to properly deal with residuals around median relations.

To achieve this, we closely follow the results of \citet{Bernardi13} and \citet{Meert13}.
By fitting a series of simulations to an unbiased SDSS galaxy sample, \citet{Meert13}
found that single S\'{e}rsic models of SDSS data are usually recovered with a precision of
0.05-0.10 mag and 5-10 per cent in radius.
\citet{Bernardi13} then pointed out that significant systematics can be induced in the derivation of total
luminosities and half-light radii \re\ of massive galaxies residing in denser environments,
mainly due to issues linked to sky subtraction and exact choice of fitting light profile.
Detailed simulations \citep{Meert13} have shown that up to 0.5 magnitudes (i.e.,
0.2 dex in luminosity/stellar mass) of systematic error could affect the measurement of
the total light competing to the most massive galaxies, inducing a nearly parallel error in the estimate of \re.
The latter systematics could easily affect the final estimate of stellar mass at the same order of other
independent large systematics arising from, e.g., different assumptions about the stellar mass-to-light ratio.

We thus first assigned independent Gaussian \emph{statistical} errors to stellar masses
and sizes with (typical) dispersions \citep{Huertas13b} of 0.2 and $\sim 0.1$ dex, respectively
(the error in size is slightly luminosity dependent following \citealt{Bernardi13},
but simply keeping it constant to 0.1 dex does not minimally alter the results).
On top of the statistical uncertainties, to reflect the results of the simulations
of \citet{Bernardi13}, we then added a \emph{systematic} variation
in predicted stellar mass. The latter is computed as follows.
We first transform each stellar mass to luminosity following the mass-luminosity
relation of \citet{Bernardi13}. To each luminosity we then
assign the maximum possible systematic error following the
largest luminosity-dependent correction given in Fig. 1 of \citealt{Bernardi13}, which
amounts to $\Delta M_r\sim 0.014$ for $M_r \gtrsim -21$ and
progressively growing to $\Delta M_r\gtrsim 0.3$ for $M_r \sim -23$, up to $\Delta M_r\gtrsim 0.6-0.7$ for $M_r \sim -24$.
All magnitudes are then converted back to luminosities using the same mass-luminosity relation.
The corresponding size is then updated by the total change in stellar mass/luminosity as
$\Delta \log R_e \equiv \Delta \log L$.

We here note that a correlation of the type $\Delta \log R_e \sim \Delta \log L$
has also been confirmed by previous studies \citep[e.g.,][]{Saglia97,Bernardi03I}. Nevertheless,
other types of correlation error between luminosity and size may be possible.
For example, if different measurements reach to different surface brightness levels,
then one could expect a correlation closer to $\Delta \log R_e \sim 0.5\Delta \log L$.
We have verified, however, that our main results and conclusions are unaffected by the inclusion
of the latter (weaker) correlation.
We also acknowledge that, in principle, errors among observables such as stellar mass and size may not
be fully correlated.
For example, the error in stellar mass will also depend on other factors
not necessarily linked to light profile's issues,
such as the choice of templates or varying IMF \citep[see, e.g., discussion in][and references therein]{Bernardi13}.
Overall, assuming null or maximal correlation among errors will bracket the full range of possibilities.

Halo masses in the catalog by \citet{Yang07}, adopted for reference in this work,
are also maximally correlated to stellar masses via abundance matching between the total stellar mass function
of the groups and clusters, and the halo mass function of dark matter haloes.
Following \citet{Huertas13b}, in order to quantify the maximum propagated error in halo mass as a function of variation in the stellar
mass of the central galaxy, we take advantage of the full Millennium simulation.
We compute the stellar mass function of central galaxies in the \citet{Guo10} model with and without
systematic plus statistical errors, and for each case compute the corresponding median relation with halo
mass via abundance matching with the halo mass function. This provides the median, stellar mass-dependent
correction to host halo mass that must be applied to the Yang et al. catalog when varying stellar masses.
We remark that both the stellar mass function and cosmology used in \citet{Yang07}
differ a little with respect to the ones in the Millennium database, but we expect these
changes to have minimal impact in our computations given that we are here mainly interested
in the median shift in halo mass, consequent to a variation in stellar mass, not in its absolute value.

We should also point out that assigning halo masses to galaxies via abundance matching
could alter both the slope and intrinsic scatter in the true stellar mass-halo mass relation.
Nevertheless, \citet{Huertas13b} proved that environmental dependence in the local Universe is not present
even when adopting \emph{fully independent} cluster mass measurements.
In the specific, they showed that field massive galaxies
share very similar size distributions to equally massive counterparts residing in
88 low-$z$ clusters from \citet{Aguerri07}, with dynamically mass measurements obtained
from the velocity distribution of spectroscopically confirmed galaxy members.
We will thus present and discuss Monte Carlo simulation results in which we also allow the error on host halo mass to be fully
uncorrelated to other quantities (though continuing to force maximal correlation between size and stellar mass).

\subsection{Broad halo distributions at fixed stellar mass}
\label{subsec|connectmhalo}

Before discussing sizes and their connection to environment,
it is clearly instructive to investigate the predictions of
models with respect to the distributions of stellar masses in different environments,
that hereby we physically identify with host group/cluster halo masses.

We first show in \figu\ref{fig|PMhaloFixedMstar}
the results obtained from the B06, G11, {\sc morgana}, and SHAM models
(blue/long-dashed, red/dot-dashed, violet/dotted,
and solid/orange histograms, respectively).
The left panel shows the distributions in host halo mass for central galaxies within a factor of 2 in stellar mass,
$11<\log \mstare<11.3$, while the right panel reports the distributions for galaxies with $\log \mstare/\msune>11.2$,
the latter being the actual stellar mass interval taken as reference for our study below (see \sect\ref{sec|Results}).
All models predict large distributions of halo masses
at fixed bin in stellar mass. Even for relatively narrow bins in stellar mass of a factor of two (left panel),
models redistribute galaxies along broad ranges of hosts,
differing by factors of $\gtrsim 50-100$ in halo mass.
Most models are also in broad agreement with the halo mass distributions inferred from the empirical sample
(solid, black lines), obtained, we remind, by cross-correlating the early-type galaxy
population in SDSS with the Yang et al. catalog.

\begin{figure}
    \includegraphics[width=8.5truecm]{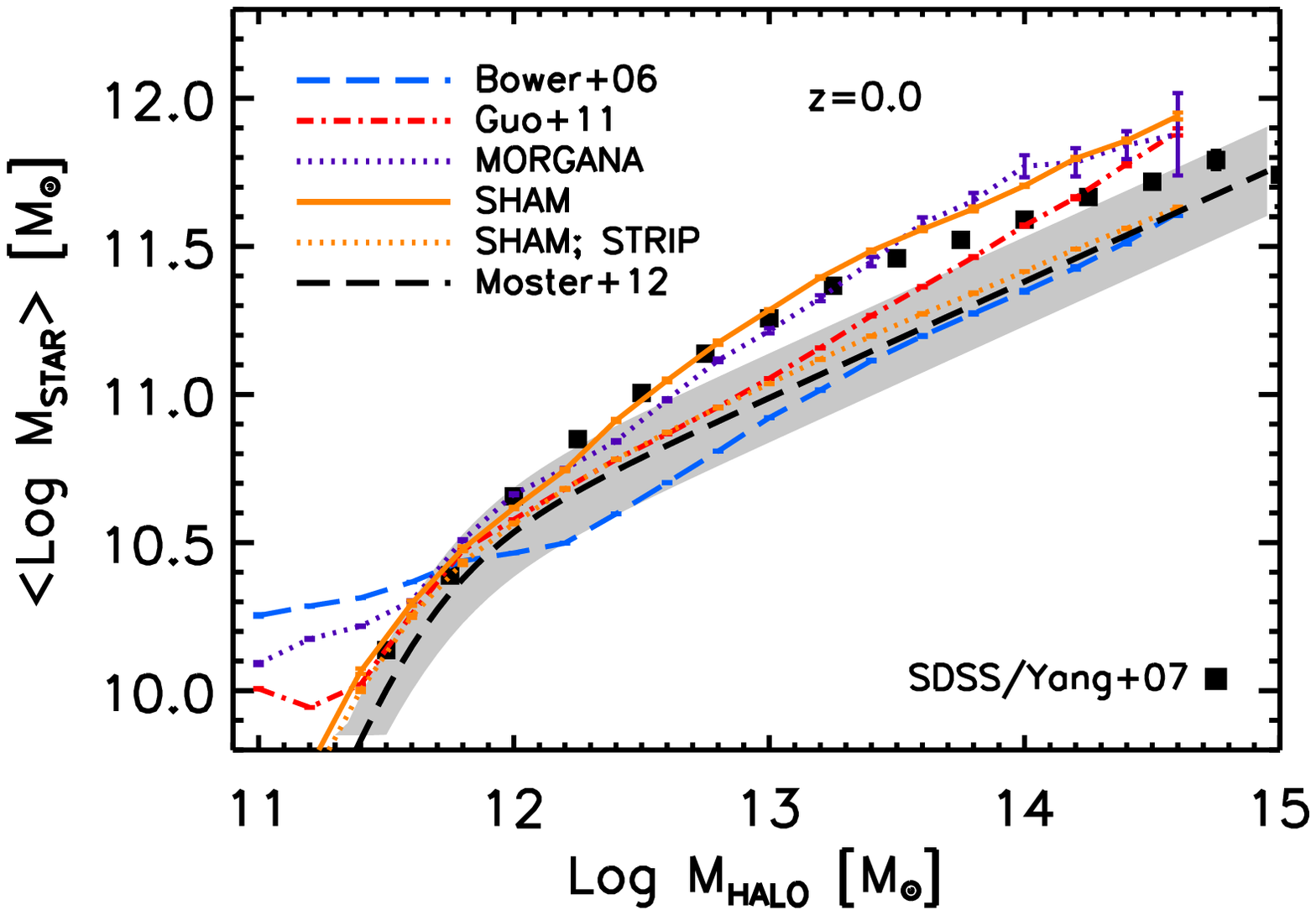}
    \caption{Predicted median stellar mass at fixed halo mass for central galaxies
    for all the reference models presented in Fig.~\ref{fig|PMhaloFixedMstar}, as labelled, at redshifts
    $z=0$. Models predict different stellar mass-halo mass relations,
    despite tuned to match the same stellar mass function, at least in the local Universe.
    The \emph{long-dashed}, \emph{black} lines are the \mstar-\mhalo\ relation derived by \citet{Moster13}
    from abundance matching techniques shown along with their 1-$\sigma$ dispersion (\emph{grey bands}).
    The \emph{solid squares} are the median \mstar-\mhalo\ relation
    competing to central SDSS early-type galaxies, derived from the \citet{Huertas13b} galaxy catalog
    matched to the \citet{Yang07} halo catalog (see text for details).}
    \label{fig|MstarMhaloRel}
\end{figure}

The analysis in \figu\ref{fig|PMhaloFixedMstar}
is restricted  to central galaxies only, but satellites
cover even broader halo mass distributions.
More generally, we find the broadening to be independent of the exact stellar mass bin considered,
or the exact \BT\ threshold chosen (here we set \BT$>0.5$),  or the type of galaxy considered (i.e., central or satellite), as long
as the analysis is restricted to massive galaxies ($\log \mstare \gtrsim 10^{11}\, \msune$).
In other words, all the galaxy formation models in this work share the view that galaxies of similar stellar mass can emerge from
different environments and may have thus undergone different growth histories.
This basic feature motivates a systematic study of
galaxy structural properties at fixed stellar mass in different environments.
Despite small differences in the broadness of halo distributions at fixed stellar mass
(with the B06 predicting the largest dispersions),
all models roughly share distributions comparable to the empirical ones.
This shows that all models
predict a median halo mass at fixed bin of stellar mass in agreement
with the data. This is not
entirely unexpected, given that the models
have been tuned to broadly reproduce the local stellar mass function.

\figu\ref{fig|MstarMhaloRel} shows instead
the median stellar mass as a function of halo mass.
Due to the large scatters involved, the latter is not equivalent to
the median halo at fixed stellar mass.
It is intriguing that only two models ({\sc morgana} and SHAM) well agree with the SDSS/Yang et al. data (filled squares),
while all the others lie somewhat below in stellar mass at fixed halo mass.
The latter models are discrepant at the high mass end by a systematic factor of $\sim 2$
with the Yang et al. results, but in better agreement with $z=0$ stellar mass-halo mass
relation worked out by \citet{Moster13} via abundance matching techniques
(long-dashed line with its 1\til$\sigma$ scatter shown as a gray area).
Empirical estimates of the stellar mass-halo mass relation still in fact disagree
by a factor of a few \citep[e.g.,][]{Behroozi10,Rodriguez12},
or possibly even more according to some studies \citep[e.g.,][]{Neistein11,Yang12}, with
galaxy evolution models predicting a similar degree of discrepancy.
What is relevant to the present work is anyway exploring structural differences
in halo mass at \emph{fixed} bin of stellar mass, so factors of $\lesssim 2$ disagreement
in scaling relations are not a major limitation for the present study.

\begin{figure*}
    \hspace*{0.in}\includegraphics[width=12truecm,angle=90]{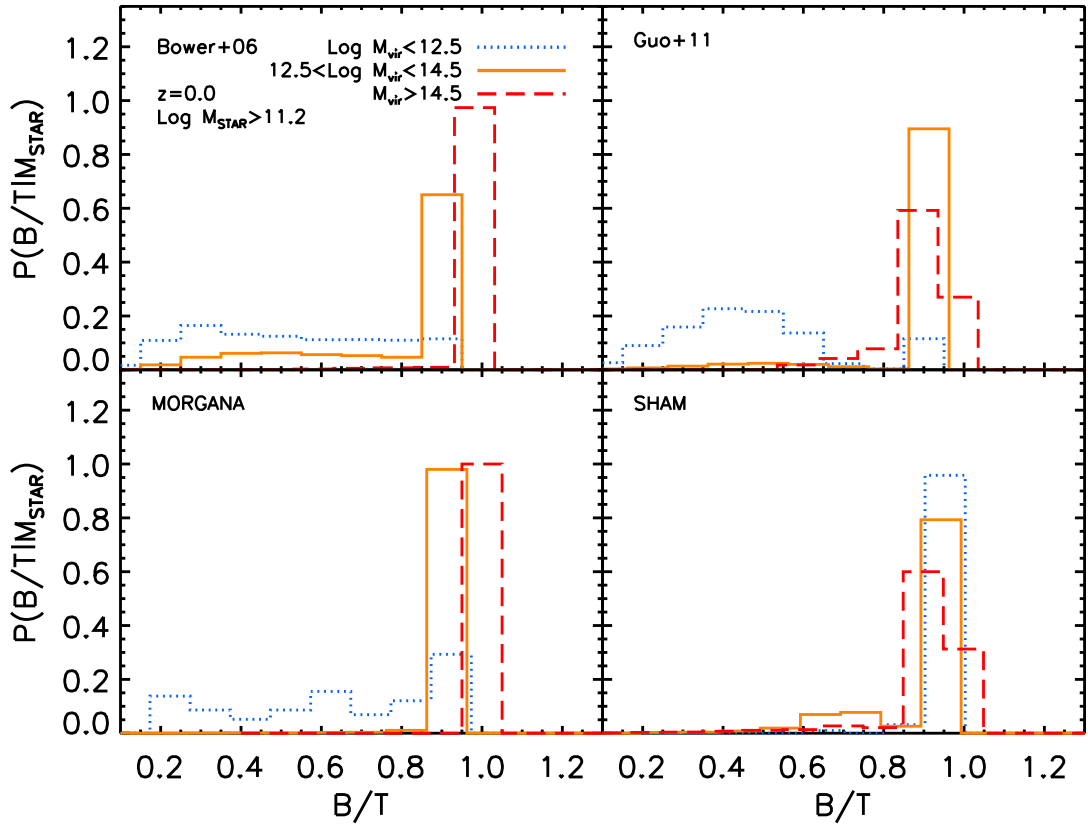}
    \caption{Predicted $B/T$ distributions for central galaxies
    with stellar mass in the range $\log \mstare/\msune >11.2$ residing in different bins of halo mass,
    $\log \mhe/\msune <12.5$, $12.5<\log \mhe/\msune >14.5$, and $\log \mhe/\msune >14.5$, for the four reference models, as labelled.}
    \label{fig|PBTFixedMstar}
\end{figure*}

We conclude the section by emphasizing that even if (proto)galaxies in the SHAM
by construction are forced to lie on the \citet{Moster13} relation (\sect\ref{subsec|SHAM}),
the resulting \mstar-\mhalo\ is predicted to be displaced upwards with respect to the $z=0$
Moster et al. relation by up to a factor of $\sim 2$ at high stellar masses.
It has already been pointed out that mergers between galaxies initialized on the
$z>0$ abundance matching \mstar-\mhalo\ relation can scatter upwards the newly
formed ellipticals \citep[e.g.,][]{Monaco06,Cattaneo11,Zavala12}. This is an effect due to the
slow evolution in the empirical \mstar-\mhalo\ relation at late times, compared to the
sudden increase in stellar mass due to mergers. The addition of significant stellar
stripping in the infalling satellites can definitely limit this tension and bring most
of the outliers back on the empirical relation at $z=0$ \citep[e.g.,][]{Monaco06,Cattaneo11}. We prove this by showing how
the version of the SHAM inclusive of stellar stripping with $\tau=0.25$ in \eq\ref{eq|stripping},
well matches the $z=0$ \citet{Moster13} relation (dotted, orange lines in \figu\ref{fig|MstarMhaloRel}).
In the following we will continue to consider the SHAM without stellar stripping (solid, orange lines)
as our reference model, as it is in better agreement with the empirical halo-galaxy catalog used as
observational constraint, although we will mention the model with stripping where relevant.

\subsection{$B/T$ distributions at fixed stellar mass bin}
\label{subsec|BTdistributions}

In order to properly compare size distributions in different
environment, it is necessary to first understand what the predictions
of the models are with respect to morphology, at least in the range of high stellar masses of interest here.

\figu\ref{fig|PBTFixedMstar} shows that all models predict quite a narrow distribution
for bulge-to-total stellar mass ratios \bt, with massive central galaxies mostly
gathered around \bt$\gtrsim 0.8$ (all distributions are normalized to unity).
Significant, long tails to the lowest values of \bt\ are however present, especially
in less massive haloes with $\mhe \lesssim 3-5\times 10^{12}\msune$.
In general, minor mergers are responsible for creating
small bulges in these models with \bt$\sim 0.1-0.3$. However, disc instabilities,
when present, inevitably drive the growth of larger bulges. The exact
resulting distribution of \bt\ is clearly dependent on the strength/type of the disc instability, and it is mainly relevant at
intermediate to low stellar masses, as also identified by previous studies \citep[e.g.,][]{DeLucia11,Shankar13}.

\begin{figure*}
    \includegraphics[width=15truecm]{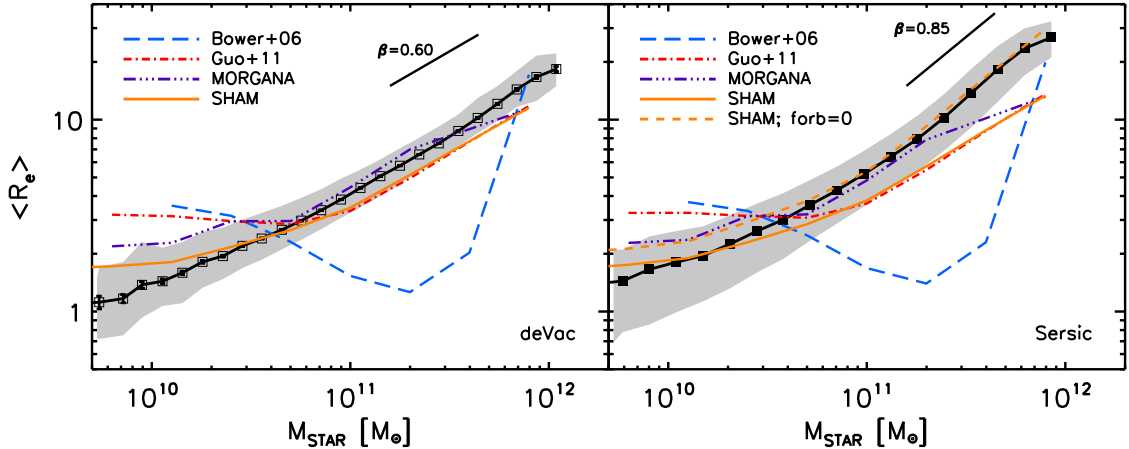}
    \caption{Predicted median size as a function of stellar mass for central galaxies
    with \bt$>0.5$ for all the models presented in \figu\ref{fig|PMhaloFixedMstar}, as labelled.
    The \emph{left panel} shows predictions converted from 3D to 2D quantities using a constant factor
    assuming $n=4$ for all galaxies, and the \emph{open squares}
    represent the relation derived by \citet{Bernardi11a} using \emph{cmodel} magnitudes.
    The \emph{right panel} shows predictions obtained by associating
    to each simulated galaxy an empirical $n$.
    In this panel we also include
    a version of the SHAM characterized by having \forb=$0$ in \eq\ref{eq|sizegeneral} (\emph{orange}, \emph{dotted} line).
    The \emph{filled squares} are the median size-stellar mass relation
    derived by \citet{Bernardi13}, using S\'{e}rsic profiles.
    The \emph{grey areas} in both panels mark the 1\til$\sigma$ uncertainty regions in the data. The S\'{e}rsic
    data are better reproduced by models having \forb=$0$.}
    \label{fig|ReMstarRel}
\end{figure*}

Although the exact shape of the \bt\ distribution characterizing each model is the result of a complex
interplay among a variety of different physical processes \citep[e.g.,][]{DeLucia11},
we can still capture some basic trends.
The B06 and {\sc morgana} models (left panels), characterized by the strongest disc instabilities,
tend to produce quite broad \bt\ distributions in the lowest mass host haloes.
This is not unexpected, as the regime where the condition in \eq\ref{eq|discinstability},
of high disc circular velocities and corresponding low host halo velocities,
is most easily met in lower mass haloes with massive galaxies.
The G11 model, with the mildest disc instabilities (cfr. \sect\ref{sec|Models}),
predicts instead more Gaussian-shaped distributions for the low-\bt\ objects, peaked around \bt$\sim 0.4-0.7$.
The SHAM, with practically negligible disc instabilities, tends to predict even lower fractions
of low \bt\ galaxies.

\figu\ref{fig|PBTFixedMstar} hints toward the fact that all hierarchical models considered in this work predict massive galaxies
in the local Universe to be bulge dominated with \bt$\gtrsim 0.8$.
More generally, we checked that all models predict a rising fraction of central galaxies with
\bt$>0.5$ as a function of stellar mass in good agreement with the data, although
models with stronger disc instabilities tend to overproduce the fraction of bulge-dominated
galaxies at stellar masses $\mstare \lesssim (1-2)\times 10^{11}\, \msune$ \citep[see also][]{Wilman13}.
What is relevant to our present discussion is that all models agree
in predicting a dominant fraction of galaxies with
high stellar masses $\gtrsim 10^{11}\, \msune$ and large
bulges built mainly via mergers. However,
most models also predict a more or less pronounced population
of galaxies with still massive bulges (\bt$\gtrsim 0.5$) residing at the centre of lower mass haloes
grown mainly via disc instabilities.
This in turn, we will see, has a non-negligible impact on the environmental dependence of sizes in lower mass haloes.

\subsection{The median size-stellar mass relation}
\label{subsec|SizeMstarRelation}

We begin our study of early-type galaxy structural properties by showing in \figu\ref{fig|ReMstarRel}
a general comparison among the median size-stellar mass relations predicted by the reference galaxy evolutionary models
against the data\footnote{Median sizes are computed from the 50\% percentile of the
full statistical distribution of galaxies competing to each
bin of halo mass considered (for the SHAM, as discussed in \sect\ref{subsec|SHAM}, the statistical
distribution is weighted by the number of effective haloes considered, although neglecting
such extra weighting makes little difference). The error on the median
is computed by dividing the 1\til$\sigma$ uncertainty of the same distribution
by the square root of the number of galaxies in that bin of halo mass.}. For the latter,
we show two estimates of the \re-\mstar\ relation.
The filled squares (right panel) represent the median size-mass relation from the data discussed
in \sect\ref{sec|Data}. Sizes are based on S\'{e}rsic profiles
\citep{Sersic63} taken from \citet{Bernardi13}. The open squares (left panel)
represent instead the relation derived for a SDSS sample of early-type galaxies
by \citet{Bernardi11a} using \emph{cmodel} magnitudes, a combination
of a de Vaucoulers \citep{deVac} and exponential profiles, as discussed in \citet{Bernardi10}.
The latter relation was calibrated on a sample of early-type galaxies with no restriction on centrals.
More generally, the sample used by \citet{Bernardi11a} is not
exactly matched to the one in the right panel, but this is irrelevant to our present discussion.
We here present both relations compared to models to simply emphasize the typical systematic
observational uncertainties that inevitably affect size measurements.
As anticipated in \sect\ref{subsec|comparisonstrategy}, fitting galaxy with
different model profiles can yield different sizes at fixed luminosity/stellar
mass up to a systematic variation of $\gtrsim 50\%$,
as seen in \figu\ref{fig|ReMstarRel}, when comparing
de Vaucouleurs (left panel) and S\'{e}rsic (right panel) profiles.
For the rest of the paper we will refer only to sizes derived from S\'{e}rsic profiles,
as they are a better fit to the light profiles of massive galaxies \citep[][and references therein]{Bernardi13}.

As in S13 when comparing model 3D half-mass radii \rh\ to measured 2D projected half-light radii \re,
we assume that light traces mass and convert \rh\ to \re\ using the
tabulated factors from \citet{Prugniel97}, i.e., $R_e\approx 2 S(n) R_H$, with the scaling factors $S(n)$
dependent on the S\'{e}rsic index $n$.
Our mock galaxy catalogues lack predictions on the evolution of $n$ competing to each galaxy.
In principle, it is possible to predict a S\'{e}rsic index a priori from the models
\citep[e.g.,][]{Hop08FP}, but this relies on several additional assumptions
on the exact profile and its evolution with time of the dissipational and dissipationless components,
that the true advantage with respect to simply empirically assign a S\'{e}rsic index at $z=0$ is modest.
We thus prefer to stick to a minimal approach with the least set of physical
assumptions and corresponding number of parameters.

We first compute 3D sizes from energy conservation arguments, as detailed in \sect\ref{sec|Models}.
Given that we are here mainly interested in bulge-dominated massive galaxies, we could simply set
$n=4$ (i.e., $S(4)=0.34$ from Table 4 of \citealt{Prugniel97}), an average value characterizing
such galaxies \citep[e.g.,][and references therein]{Huertas13a,Bernardi13}.
The left panel of \figu\ref{fig|ReMstarRel} shows predictions converted from
3D to 2D quantities using a constant $n=4$ for all galaxies.
The right panel of \figu\ref{fig|ReMstarRel} shows instead
predictions obtained via a mass-dependent $n$, in which each mock galaxy
has been assigned an index $n$ from our empirical
$n$-$\log \mstare$ relation (\sect\ref{sec|Data}).
We find the outputs to be so similar that including or not a mass-dependent
conversion factor $S(n)$ makes little different to our results below.
For consistency with the data, in the following we will continue
adopting the mass-dependent S\'{e}rsic correction as our reference one.

There are several general noteworthy features in \figu\ref{fig|ReMstarRel}.
First, models, despite the different details in computing galaxy
stellar masses and sizes, predict quite similar \re-\mstar\ relations,
both in shape and normalization, especially for $\mstare \gtrsim 3\times 10^{10}\, \msune$.
The broad agreement with the data is also reasonable, most models
lie within the 1\til$\sigma$ uncertainties of sizes at fixed stellar mass
(dotted lines), except for the B06 model, which significantly diverges from them.
This has been extensively discussed by \citet{Gonzalez09} and \citet{ShankarPhire}, and it can
be ascribed to some possibly wrong initial conditions, as a similar
behavior is present also at higher redshifts. In particular, varying their input
parameters does not ameliorates the match, although
an improvement can be achieved by switching off adiabatic contraction \citep{Gonzalez09}.
Despite being the B06 model highly divergent with respect to the global
size-stellar mass relation, for completeness, we will continue keeping it as one of our reference models,
even when discussing environmental dependence.

Irrespective of the exact 3D-to-2D conversion adopted, models fall short (at the $\sim$ 2\til$\sigma$ level)
in reproducing the exact normalization of the S\'{e}rsic size-mass correlation
at masses $\mstare \gtrsim 1-2\times 10^{11}\, \msune$ (right panel), indicative of a more profound cause of discrepancy.
The short-dashed, orange line shows the prediction of the SHAM model with the same specifications as the reference
one but with \forb$=0$ (i.e., negligible orbital energies, parabolic orbits).
As anticipated by S13, this variation in the merging model effectively produces larger sizes for the
same stellar mass because each merger event is more efficient in enlarging the central galaxy (\eq\ref{eq|sizegeneral}).
More massive galaxies which are the most affected by mergers, will
proportionally be larger resulting in an overall steepening of the size-stellar
mass relation and a significantly better agreement with the data (cfr. solid/orange and short-dashed/orange lines).
The predicted slope of the size-stellar mass relation at high masses steepens from $\beta\sim 0.6$ (left panel)
to $\beta\sim 0.8-0.9$ (right panel) when setting \forb$=0$.

\begin{figure}
    \includegraphics[width=8.5truecm]{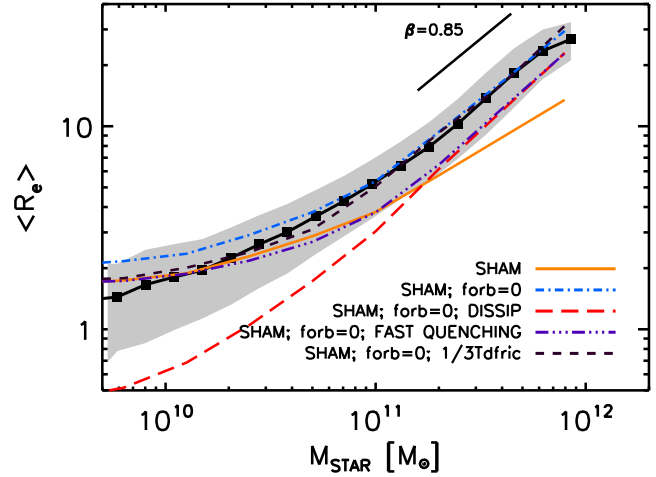}
    \caption{Predicted median size-stellar mass relation for different versions of the SHAM model, as labelled,
    compared to data (\emph{solid} line with \emph{grey area} marking the 1\til$\sigma$ uncertainty region).
    All models predict similar scaling relations especially at the massive end, within a factor of $\lesssim 2$ in normalization.
    The predicted slope is around $\beta\sim 0.8-0.9$ for all models with \forb$=0$, and closer to $\beta\sim 0.60$ for models
    with \forb$>0$.}
    \label{fig|ReMstarSHAMs}
\end{figure}

In \figu\ref{fig|ReMstarSHAMs} we report the predicted size-stellar mass relation for the different SHAMs
introduced in \sect\ref{subsubsec|SHAMmodels}, with null input median orbital energy,
with gas dissipation, with fast satellite quenching, and with a shorter dynamical friction timescale.
Analogously to what emphasized with respect to \figu\ref{fig|ReMstarRel},
SHAM models with \forb$=0$ share a slope at the massive end of $\beta\sim 0.8-0.9$, in better
agreement with the observed one, while $\beta\sim 0.6$ for models with \forb$>0$.
While for consistency with the other models, we will continue using \forb$>0$ in the SHAM
reference model, we will extensively discuss models characterized by \forb$=0$ which represent a better match to the S\'{e}rsic data.

Irrespective of details on the exact galaxy profile, the different steepening
in the size-stellar mass relation for different models could be qualitatively probed directly from \eq\ref{eq|sizegeneral}.
As already sketched several times in the literature \citep[e.g.,][and references therein]{Bernardi09,Naab09,ShankarBernardi09},
we can in fact write the increase in radius due to mergers as
\begin{equation}
\frac{R_{\rm new}}{R_1}=\frac{(1+f)^2}{(1+f^2/\eta+kf/(1+\eta))}
\label{eq|RnewApprox}
\end{equation}
with $f=M_2/M_1$, $\eta=R_2/R_1$, and $k=f_{\rm orb}/c$. For simplicity, considering a merger history dominated
by (very) minor mergers with $f^2<<f<<1$, and $\eta \approx f$ we can set, after some approximations,
\begin{equation}
\beta=\frac{\Delta \log R}{\Delta \log \mstare}=\frac{\Delta R}{R}\frac{\mstare}{\Delta \mstare}\approx
\frac{f(1-k)}{(1+kf)}\frac{1}{f} \nonumber
\label{eq|BetaApprox}
\end{equation}
which yields $\beta \gtrsim 0, 1$ for $k \sim 1, 0$ respectively. Clearly the latter approximations
are very basic and cannot capture the full complexities behind galaxy merger histories, but nevertheless clearly highlight
how the slope of the size-stellar mass relation can easily steepen for lower values of \forbcl.
In other words, the slope of the size-stellar mass relation at high masses in hierarchical models is
more a consequence of the \emph{type} rather than the number of galaxy mergers.

It is evident from \figus\ref{fig|ReMstarRel} and \ref{fig|ReMstarSHAMs},
that despite the different input physical assumptions,
most of the models predict similar size-stellar mass relations within the $1\til\sigma$ uncertainties of the data (grey bands).
The differences are within a factor of $\lesssim 2$ at high stellar
masses $\gtrsim 10^{11} \msune$, the ones of interest here.
The predicted behaviour among different models
below \mstar$\lesssim 10^{11} \msune$ is instead somewhat varied.
Most of the semi-analytic models in \figu\ref{fig|ReMstarRel}
predict a more or less pronounced flattening at lower masses, while the SHAMs
in \figu\ref{fig|ReMstarSHAMs} tend to mostly align with the data, except
for the SHAM with gas dissipation which tends to progressively fall below the data
at low stellar masses.
S13 discussed that the low mass end shape of the resulting
size-stellar mass relation depends, among other factors, on the exact slope of the underlying \mstar-\mhalo\ relation
in the same stellar mass range, thus explaining part of the discrepancies among different models.
Furthermore, as discussed by \citet{Hop08FP} and \citet{Covington11},
gas dissipation can effectively shrink the sizes of lower mass bulges, remnants
of gas-richer progenitors. S13 showed that this mechanism
can indeed significantly ameliorate the match to the data in the G11 model, entirely removing the flattening
in sizes at low masses.
On the other hand, the SHAM with gas dissipation tends to drop at low masses more rapidly than the G11 model
with gas dissipation (see full discussion and related Figures in S13), possibly due to the different
\mstar-\mhalo\ relations, input gas fractions, and detailed treatment of satellites.

\begin{figure*}
    \hspace*{0.in}\includegraphics[width=12truecm,angle=90]{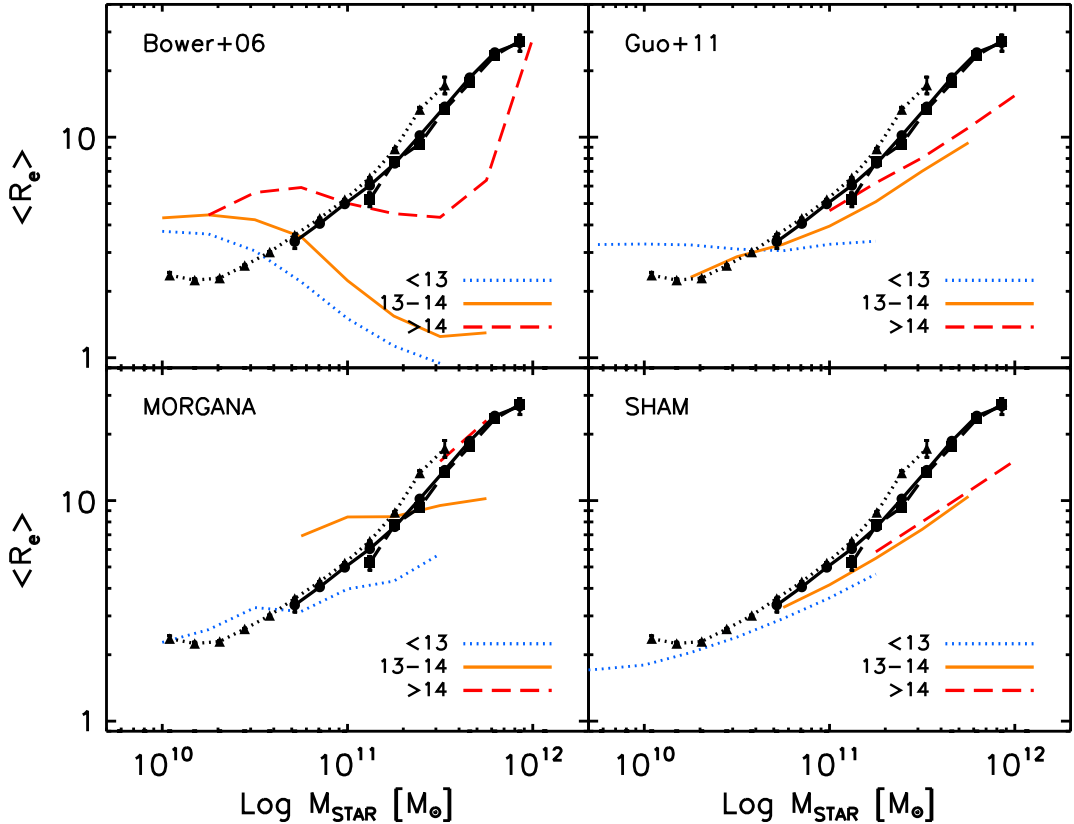}
    \caption{Predicted median size-stellar mass relation of central galaxies in different bins of halo masses,
    $\log \mhaloe/\msune<13$, $13<\log \mhaloe/\msune<14$, and $\log \mhaloe/\msune>14$, for different models, as labelled.
    Data (black lines with symbols) are as in \figu\ref{fig|ReMstarRel}, divided by the same
    bins of host halo mass as in the models. All models present a more or less pronounced
    variation up to a factor of $\lesssim 3$ of median size at fixed
    stellar mass, when moving from low to high mass haloes.}
    \label{fig|ReMstarFixedMhalo}
\end{figure*}

Some properly fine-tuned disc regrowth/survival
after a gas-rich major merger \citep[e.g.,][]{Puech12,Zavala12}
could boost the total sizes of low mass galaxies, thus improving the
match between the data and the SHAM with gas dissipation (but then worsening the good one with the G11/S13 model).
\citet{Bernardi13} have indeed recently stressed that the contribution
of a disc component in early-type samples becomes
increasingly more important below $\lesssim 10^{11}\, \msune$,
while the size of the bulge component becomes
progressively more compact. The latter may then require on one side
gas dissipation to get enough compact bulges \citep[see also discussion in][]{Hop08FP}, and
on the other possibly some properly fine-tuned disc regrowth \citep[e.g.,][]{Puech12}
to recover the disc components measured in these galaxies.
A full treatment of the general impact of gas dissipation and/or disc regrowth
models in low mass galaxies is beyond the scope of the present work,
and in the following we will mainly focus on the impact of gas dissipation
alone in the context of environmental dependence of very massive spheroids.

To summarize, \figus\ref{fig|ReMstarRel} and \ref{fig|ReMstarSHAMs} prove
how the size-stellar mass relation by itself may not represent a major discriminant for determining
the success of a galaxy evolution model with respect to another, especially for
bulge-dominated galaxies above $\gtrsim 10^{11}\, \msune$.
In the following, we will discuss how galaxy sizes coupled with the notion on
their environment, defined here as the host total halo mass,
can provide useful additional physical insights.

\subsection{Environmental Trends}
\label{subsec|trend}

To start off with, \figu\ref{fig|ReMstarFixedMhalo}
reports for both models and data, the
median sizes competing to galaxies of similar stellar mass but
living in different environments, in the specific,
at the centre of haloes of mass $\log \mhaloe/\msune<13$, $13<\log \mhaloe/\msune<14$, and $\log \mhaloe/\msune >14$
(marked by dotted, solid, and long dashed lines, respectively).
Only bins with at least 20 galaxies are retained in this plot.

While the observed size-stellar mass relation seems to be quite ubiquitous in all environments,
there is a net tendency for models to predict larger galaxies
in more massive haloes.
This tendency is marginal for the models on the right panels,
while significantly more pronounced in the models reported in the left panels.
In the specific, the reference SHAM predicts a rather moderate environmental
dependence of up to $\sim 30\%$ at fixed stellar mass,
the G11 model up to a factor of $\sim 2$, the {\sc morgana}
up to $\lesssim 3$, and the B06 model up to even a factor of $\lesssim 5$ in the most massive bins.
The latter two models, we recall, are the models characterized by the strongest disc instabilities (cfr. \sect\ref{sec|Models}),
thus possibly suggesting that this physical process may contribute to such a trend.

\begin{figure*}
    \hspace*{0.in}\includegraphics[width=12truecm,angle=90]{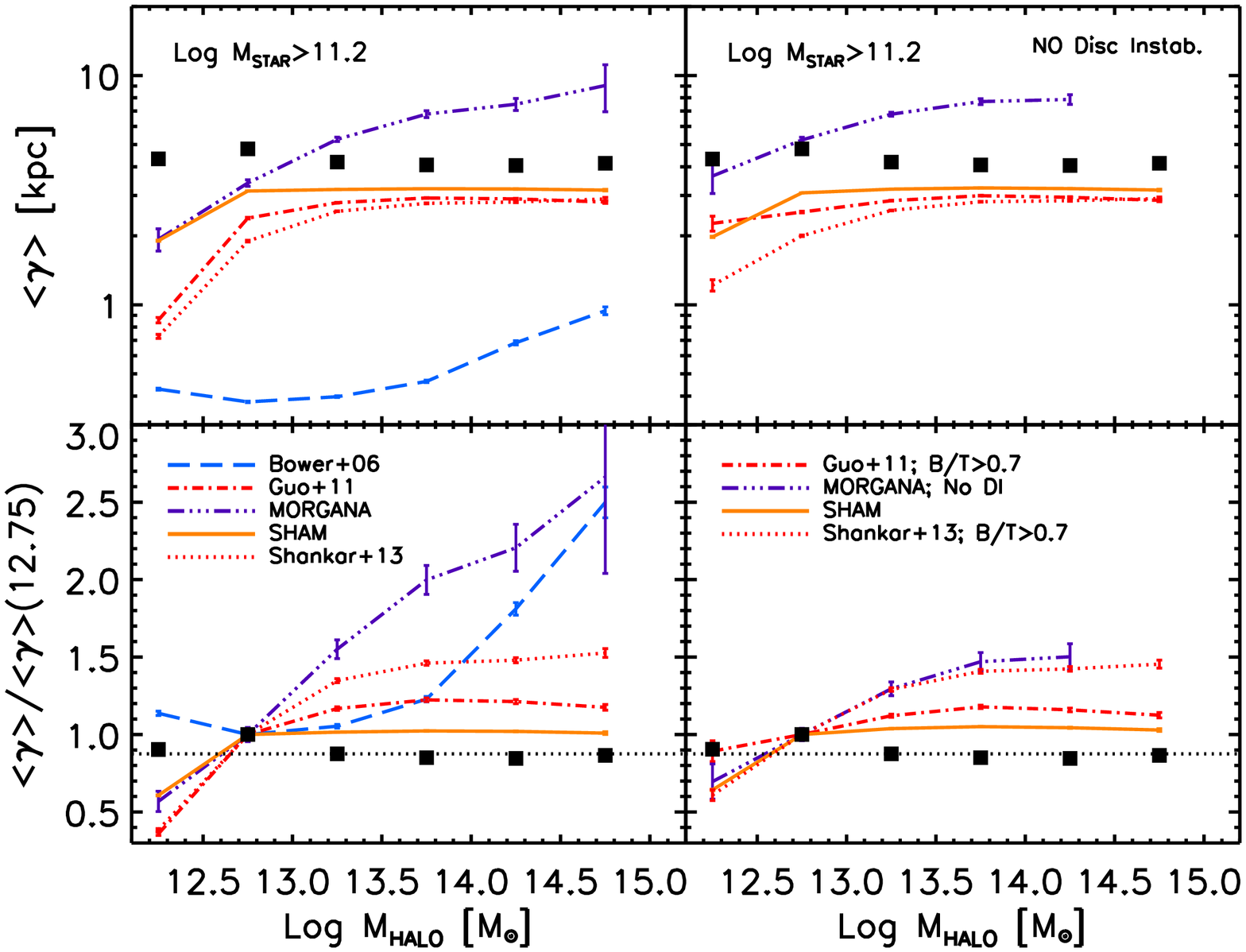}
    \caption{Fractional increase of median size in the stellar mass bin $\log \mstare/\msune >11.2$
    and \bt$>0.5$ for the different type of reference models discussed in \sect\ref{sec|Models}, as labelled.
    Data have been discussed in \sect\ref{sec|Data}. Models with weak or absent disc instabilities are favoured.}
    \label{fig|ReMstarFixedMhaloNorm}
\end{figure*}

\subsubsection{What is driving the trend?}
\label{subsec|modeltrend}

As extensively discussed in \sect\ref{subsec|SizeMstarRelation},
most of the models considered in this work align within the 1\til$\sigma$
uncertainty with the high-mass end of the local size-stellar mass relation.
Moreover, a simple change in one of the parameters such as, e.g., \forb,
can help to further fine-tune the models to match the data.
Thus, both the slope and normalization of the \re-\mstar\ relation cannot really
be effective in distinguish among the successful models.
On the other hand, the residuals around the median relation
can provide useful additional hints to constrain the models, as proven below.

Here we re-propose the same argument of \figu\ref{fig|ReMstarFixedMhalo}
but in a different format. Following, e.g., \citet{Cimatti12,Newman12},
we first select galaxies in a given bin of stellar mass in the range $M_1$ and $M_2$ and then normalize their sizes following
the relation
\begin{equation}
\log \gamma=\log R_e+\beta(11-\log \mstare) \, .
\label{eq|Renorm}
\end{equation}
\eq\ref{eq|Renorm} allows to weight each size by its appropriate stellar mass,
according to its (median) position on the size-mass relation. This way galaxies within the bin
which appear larger/smaller because more/less massive,
are properly renormalized removing any spurious effect in the study of residuals around the relation.

The slope $\beta$ in \eq\ref{eq|Renorm} is then for each model self-consistently computed
in the range of stellar mass $11.2<\log \mstare/\msune<12$, the one
of interest in this work (\sect\ref{subsec|comparisonstrategy}).
As shown in \figu\ref{fig|ReMstarSHAMs}, most of the models characterized by \forb$=0$,
have a slope of $\beta \gtrsim 0.8$ in this mass range,
in close agreement with the high mass-end slope present in the data.
Models which also include gas dissipation in major mergers, e.g., one version of the SHAM and S13,
are the ones characterized by the steepest relations at the massive end with $\beta \sim 0.9$
(cfr. \figu\ref{fig|ReMstarSHAMs}). All other galaxy models characterized by \forb$>0$,
tend to have a shallower slope of $\beta \sim 0.60$ (see \figu\ref{fig|ReMstarRel}).

\figu\ref{fig|ReMstarFixedMhaloNorm} shows the predicted mass-normalized sizes \gammar\
as a function of host halo mass for all central galaxies with \bt$>0.5$.
Both the left and right panels comprise
the outputs from the compilation of our reference models, the
B06, G11, {\sc morgana}, and basic SHAM, as labelled.
For completeness, we also add the predictions of the S13 model, introduced in \sect\ref{subsec|G11model}, variation
of the G11 model which, we remind, includes gas dissipation in major mergers and a value of \forb$=0$.

To further highlight the true information on the residuals,
we normalize each \gammar\ to one single value,
thus removing the effect of the global median normalization in the \re-\mstar\ relation.
In the lower panels \figu\ref{fig|ReMstarFixedMhaloNorm}, median \gammar\ sizes
have been divided\footnote{We choose to normalize in the interval $12.5<\log \mhaloe/\msune<13.0$ as this is the lowest mass bin
in halo mass retaining a significant number of massive galaxies (see \figu\ref{fig|PMhaloFixedMstar}).}
by the median \gammar\ competing to galaxies residing in haloes
with mass in the range $12.5<\log \mhaloe/\msune<13.0$,
to emphasize any difference in median size when moving from lower to higher mass haloes hosting central galaxies
of the same stellar mass (which could
in principle be induced by either larger galaxies at the centre of clusters, and/or more compact galaxies in the field).

\begin{figure*}
    \hspace*{0.in}\includegraphics[width=12truecm,angle=90]{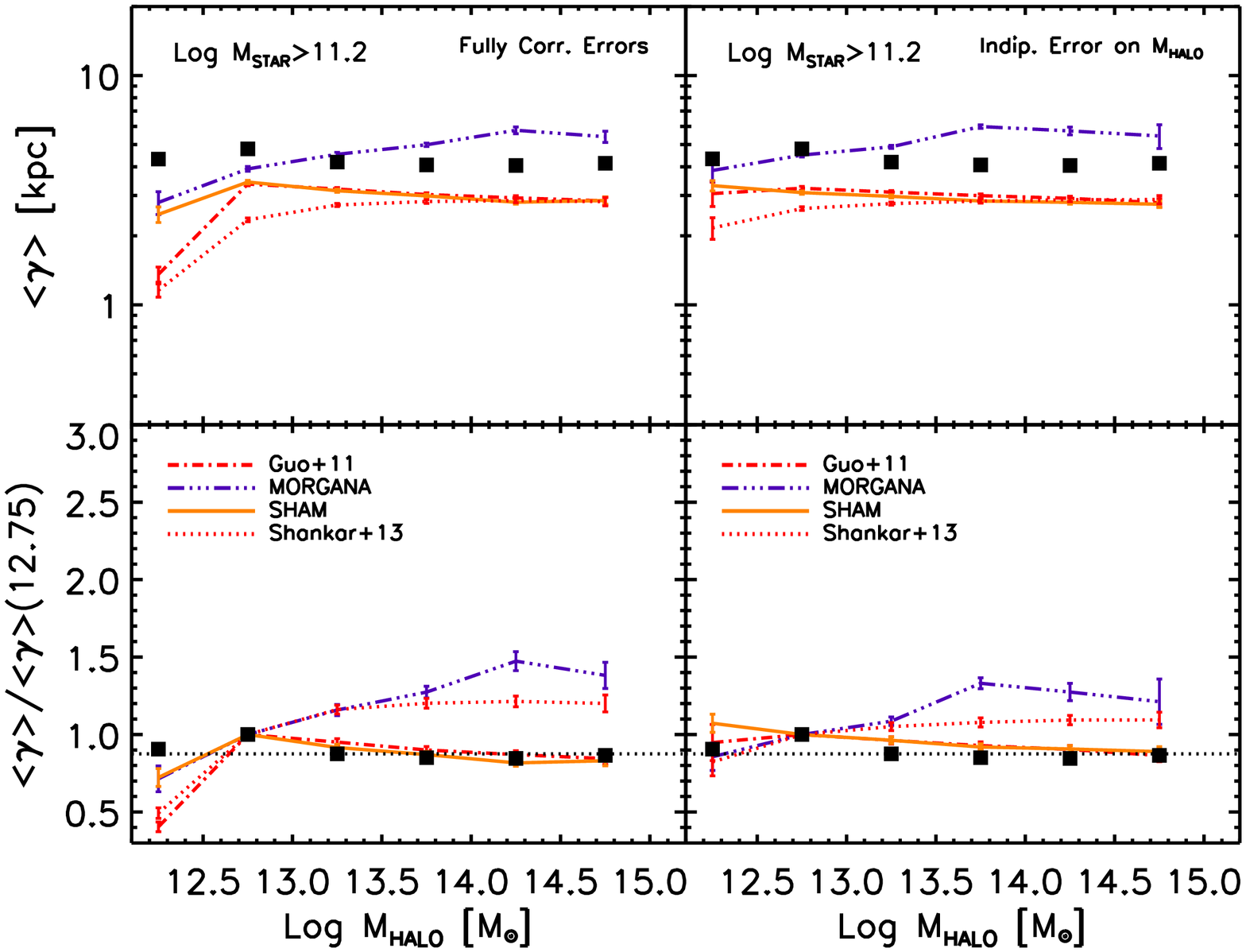}
    \caption{\emph{Top}: Predicted median \gammar\ sizes
    for the same set of of models as in \figu\ref{fig|ReMstarFixedMhaloNorm}, as labelled. \emph{Bottom}: Corresponding
    fractional increase of median size. In the \emph{left} panels
    all predictions are convolved with fully correlated errors, while in the \emph{right} panels
    we assume errors in halo mass to be independent (see text for details).
    Models with strong gas dissipation in major mergers, and/or very low dynamical
    friction timescales are disfavoured, although the effect is weakened
    in the presence of specific combinations of correlated errors.}
    \label{fig|ReMstarFixedMhaloNormMore}
\end{figure*}

In the left panels a fast variation in the median \gammar\
by a factor of $\sim 1.5-3$ is evident in most galaxy evolution models,
when moving from field/groups to cluster scale host haloes.
This behaviour is clearly at variance with the data
which suggest a flat size distribution as a function of halo mass, as indicated by the horizontal, dotted
line which marks the average normalized \gammar\ value in the data.
The discrepancy between model predictions and data is at face value highly significant.
The reference SHAM is the only one predicting a very mild variation, up to a factor of
$\sim 1.3-1.5$ or so, and nearly absent above haloes of mass $\log \mhaloe/\msune \gtrsim 13$.
We will further discuss variations to the reference SHAM below.
Here we highlight that models characterized by mergers, strong disc instabilities (B06 and {\sc morgana}),
and/or significant gas dissipation in (major) mergers (S13) predict, on the contrary, large discrepancies with the data.

To isolate the role of mergers with respect to that of disc instabilities, the right panels
of \figu\ref{fig|ReMstarFixedMhaloNorm}
show the same models but with null or minimal contribution from disc instabilities.
To this purpose, we restrict the predictions of the G11 and S13 models
to the subsamples of galaxies with
\bt$>0.7$, a limit above which it was shown bulges grow mainly via mergers \citep[see discussion in][]{Shankar13}.
In the same panels we also report a variation of the {\sc morgana} model
\emph{without} any disc instabilities. We keep for reference
the SHAM model, for which the contribution of disc instabilities is already negligible, as anticipated in \sect\ref{subsec|SHAM}.

It is interesting to note that in the absence of disc instabilities
the environmental dependence is reduced in all models, in the sense that
galaxies living in host haloes of mass $\log \mhaloe/\msune \lesssim 13$
tend to be larger than galaxies of comparable stellar mass and in the same haloes but lower \bt,
while median sizes remain less affected beyond this halo mass scale.
This behaviour is directly explained by the fact
that disc instabilities (\eq\ref{eq|discinstability}), most frequent
in lower host dark matter haloes, are less efficient than mergers in producing large bulges
of comparable mass, as anticipated in \sect\ref{sec|Models}
(cfr. \eq\ref{eq|sizegeneral} and \eq\ref{eq|sizediscinstab}).
Overall, models in which bulges significantly grow via impulsive and exceptionally
strong disc instabilities will inevitably, in these models, produce smaller bulges, preferentially
in lower mass haloes, thus enhancing any environmental dependence.

\begin{figure*}
    \hspace*{0.in}\includegraphics[width=12truecm,angle=90]{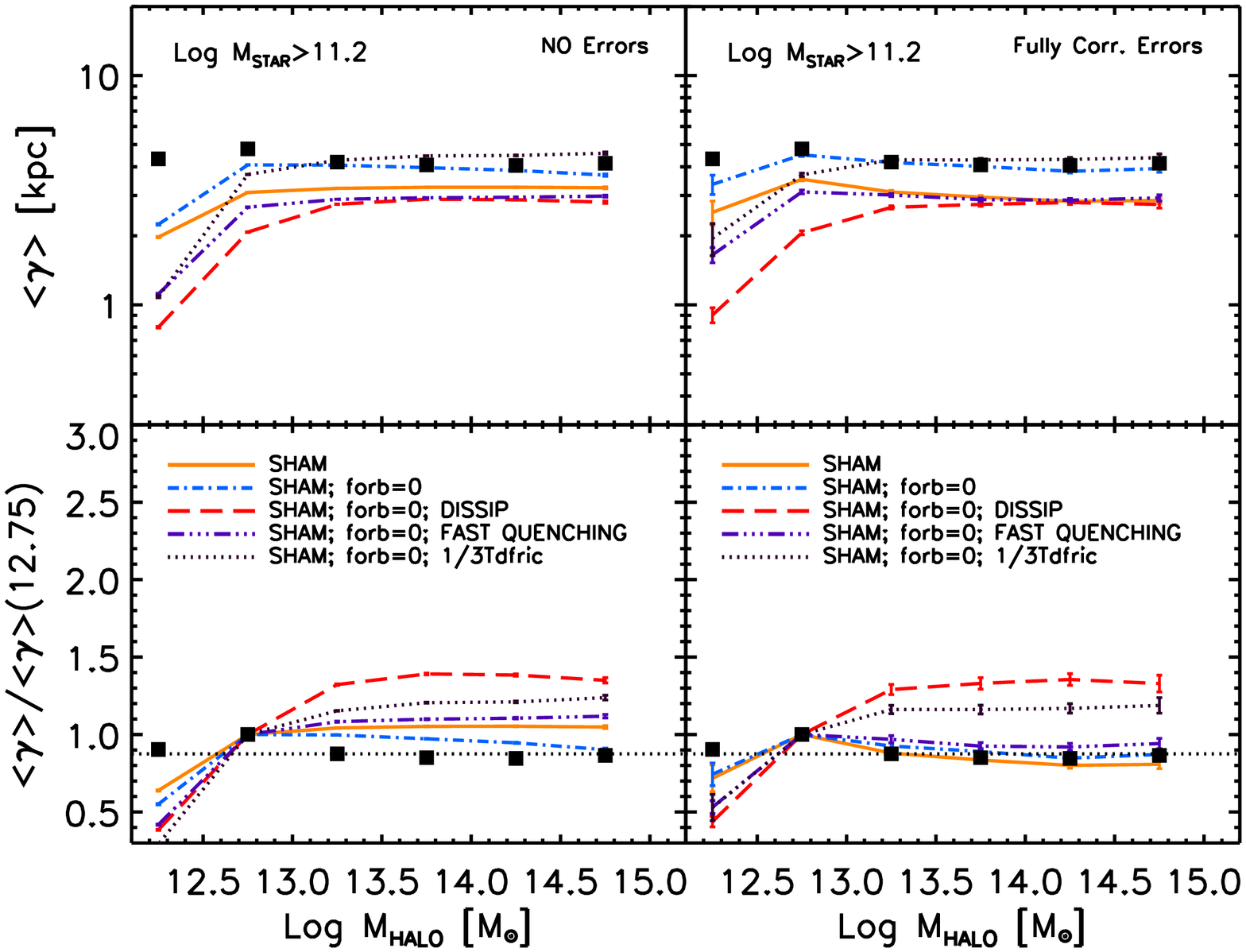}
    \caption{Same format as \figu\ref{fig|ReMstarFixedMhaloNorm}
    for different SHAM models, as labelled. The \emph{left} panels
    report raw predictions, the \emph{right} panels
    present models predictions after convolution with errors. Even in the semi-empirical formalism,
    models with strong gas dissipation in major mergers, and/or very low dynamical friction timescales, tend to be disfavoured.}
    \label{fig|ReMstarFixedMhaloNormSHAM}
\end{figure*}

Nevertheless, even in the absence of disc instabilities (right panels), most models continue
to predict a factor of $\sim 1.3-2$ differences in the median sizes as a function of halo mass,
which implies that other physical processes are contributing to environmental dependence.
The G11 model shows an increase in median size by a cumulative factor of $\sim 1.4$,
and the {\sc morgana} model with no disc instabilities predicts even more.
The latter effect may be due to the fact that in the {\sc morgana} model
bulge growth via minor mergers is more efficient than in the G11 one,
as in the former the whole baryonic mass of the satellite is transferred to the bulge of the central (\sect\ref{sec|Models}).
The S13 model also shows stronger environmental dependence with respect to the original G11 model,
thus implying that the inclusion of gas dissipation and/or the null value of \forb\ can contribute to this increase.
We will further dissect the role played by different processes,
by making use of the variations to the reference SHAM model
introduced in \sect\ref{subsec|varyingSHAMs}.

\subsubsection{A closer comparison to the data: Including observational errors}
\label{subsec|comparedata}

As anticipated in \sect\ref{subsec|comparisonstrategy}, a proper
comparison to the data requires convolution with observational errors.
To achieve this goal, we follow the methodology outlined in \citet{Huertas13b}.
In each bin of halo mass, we randomly extract from the mocks
a number of galaxies equal to the number actually extracted from the SDSS/Yang et al. catalog.
We then include errors in all the variables of interest
following the methodology outlined in \sect\ref{subsec|comparisonstrategy},
recalibrate the slope $\beta$ of the size-stellar mass
relation, and finally recompute mass-dependent sizes following \eq\ref{eq|Renorm}.
We repeat the above process for 1000 times and for each mock realization compute median sizes.
From the final distributions of medians we extract the final median value and its 1\til$\sigma$ uncertainty.
When dealing with galaxy populations in groups and cluster environments one should also consider
field contamination, as we did in previous work \citep{Huertas13b}.
However, we neglect the latter effect as we are here mainly interested in central galaxies.

We found that simply including only independent errors in all the three variables,
namely size, stellar mass, and halo mass, does not really alter the raw model
predictions presented in the previous sections.
This is mainly due to the fact that reasonable errors in stellar mass ($\gtrsim 0.2$ dex) tend to
preserve or even boost trends of median size with environment at fixed stellar mass.
Lower mass and more compact galaxies preferentially residing in lower mass haloes,
enter the selection creating a spurious increase of environmental dependence, or enhancing any pre-existing one
\citep[see simulations in][]{Huertas13b}.

More interesting to our purposes is instead the case of maximally correlated errors in size and stellar mass, and this is the one
which will be discussed in this section. In the latter scenario,
we found in fact that this combination of errors can produce an effective reduction of the environmental signal, thus providing
a viable possibility to better reconcile model predictions with observational results.
We checked that, as expected, fully correlated errors in size and stellar mass, while possibly relevant for
environmental trends, do not significantly
alter the slope $\beta$ of the intrinsic size-stellar mass correlations, thus fully preserving the
results discussed in \sect\ref{subsec|SizeMstarRelation}.
This is expected as varying size and stellar mass in a correlated way, tends
to preferentially move galaxies along the relation. The total scatter, however,
tends to somewhat increase up to about $\lesssim 30\%$,
irrespective of the exact model. We will discuss the relevance of this effect to our general discussion
in \sect\ref{subsec|ScatterReMstarAndBTfractions}.

The left panels of \figu\ref{fig|ReMstarFixedMhaloNormMore} report the results of our Monte Carlo simulations
including fully correlated errors in all the three variables. As for \figu\ref{fig|ReMstarFixedMhaloNorm},
the top panels report the median \gammar\ competing to galaxies within the chosen interval of stellar mass ($\log \mstare/\msune >11.2$),
while the lower panels show the same curves normalized to the value in haloes $12.5<\log \mhaloe/\msune<13$.
By comparing with the left panels of \figu\ref{fig|ReMstarFixedMhaloNorm},
it can be seen that the inclusion of proper errors can alter the raw model outputs by significantly reducing the
increase in median size with halo mass. In particular, we find that
models predicting $\lesssim 30-40\%$ of environmental dependence (e.g., SHAM and G10),
tend to be flattened out after inclusion of correlated errors.
On the other hand, models with stronger environmental dependence at the level of
$\gtrsim 50\%$ increase in median size when moving from field to clusters (e.g., S13 or {\sc morgana}),
tend to preserve significant size segregation. These findings
are consistent with the results of the independent and different Monte Carlo tests performed by \citet{Huertas13b}.

In the right panels of \figu\ref{fig|ReMstarFixedMhaloNormMore} we show predictions for the
same models as in the left panels with errors in halo mass this time uncorrelated to the other
quantities, following a Gaussian with dispersion of 0.3 dex (see \sect\ref{subsec|comparisonstrategy}).
It is clear that in this case any trend with environment is further suppressed by up to an extra factor of $\lesssim 1.5$, leaving
at most a factor of $\lesssim 20\%$ at the highest halo masses.
At variance with errors in stellar masses,
substantial independent errors in halo mass tend to naturally further mix the host halo masses
of galaxies at fixed stellar mass, thus contributing to reduce any signal with environment.

In conclusion, we showed that when including correlated errors,
especially in size and stellar mass, can alter model predictions and dump a significant part of the
signal with environment. Estimating exact observational uncertainties, and correlations among them,
becomes thus fundamental to break degeneracies in the models.

\subsubsection{Varying the SHAM reference model}
\label{subsec|varyingSHAMs}

As evident from \figus\ref{fig|ReMstarFixedMhaloNorm} and \ref{fig|ReMstarFixedMhaloNormMore},
models characterized by having strong gas dissipation (S13) and/or stronger,
impulsive bar instabilities (B06; {\sc morgana}), produce the largest discrepancies with the data.
To gain more insights into the causes of the discrepancies between hierarchical models and the data,
we discuss in \figu\ref{fig|ReMstarFixedMhaloNormSHAM} the predictions of the different variations of
the reference SHAM model introduced in \sect\ref{subsec|varyingSHAMs}, which bracket all the main physical processes discussed above.

\figu\ref{fig|ReMstarFixedMhaloNormSHAM} follows the same format as \figus\ref{fig|ReMstarFixedMhaloNorm} and
\ref{fig|ReMstarFixedMhaloNormMore},
with the left panels collecting the raw model predictions, while in the right
panels the predictions are convolved with fully correlated errors. Besides the reference SHAM (solid/orange lines),
\figu\ref{fig|ReMstarFixedMhaloNormSHAM} contains predictions for other SHAM outputs, all characterized by \forb$=0$,
a choice which better matches the local size mass relation (\sect\ref{subsec|SizeMstarRelation}).
The blue/dot-dashed lines refer to the SHAM which only varies \forb. The red/long-dashed lines refer
to the SHAM which, in addition, also includes gas dissipation in major mergers.
The magenta/triple dot-dashed lines correspond to the SHAM where satellites undergo
quick quenching after infall. Finally, the black, thick, dotted lines refer instead to the SHAM run with reduced
dynamical friction timescales.

It can be seen that models characterized by lower dynamical friction timescales, and/or strong gas dissipation
in major mergers, and/or fast quenching, tend to increase any environmental dependence, with the
gas dissipation model predicting the strongest steepening with halo mass, in line with the S13 model.
As more extensively discussed below, gas dissipation in major mergers tends to decrease the sizes of the remnants,
progressively more efficiently in lower mass haloes. This is because the satellite progenitors in these environments
tend to be relatively gas richer thus inducing more dissipation and more compact remnants.

\begin{figure*}
    \includegraphics[width=15truecm]{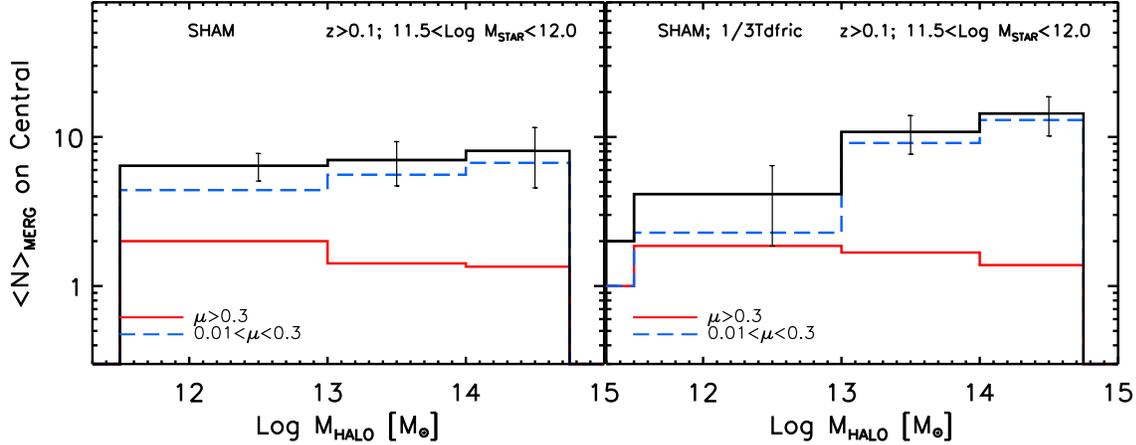}
    \caption{Predicted cumulative mean number of mergers on the central galaxy as a function of host halo mass at $z=0$.
    Marked with \emph{blue}, \emph{long-dashed} and \emph{red}, \emph{solid} lines
    are the cumulative number of minor and major mergers down to $z=0$, respectively,
    with the \emph{black}, \emph{solid} lines the sum of the two.
    The \emph{vertical bars} indicate the 1\til$\sigma$ uncertainties on the total number of mergers in each bin of halo mass.
    The \emph{left} panel shows the reference SHAM model with dynamical friction timescales \tdf\ from \citet{McCavana12}, while
    the \emph{right} panel reports the outputs of a model with 1/3 of the same merger timescales.
    Overall, it is evident that adopting the dynamical friction
    timescales expected from the analysis of high-resolution numerical simulations (\emph{left}), the number
    of mergers at fixed stellar mass does not significantly increase with halo mass.
    Here the analysis is restricted to galaxies with
    stellar masses $11.5<\log \mstare/\msune<12$ and $B/T>0.5$, but the basic result of a flattish distribution of mergers
    is similar for other selections. On the other hand, shorter dynamical friction timescales
    inevitably increase the number of (minor) mergers
    in more massive haloes, thus inducing more size growth and more environmental dependence.}
    \label{fig|Nmergers}
\end{figure*}

Just the opposite is true when lowering the merging timescale.
In the latter case lowering \tdf\ increases the number of mergers with the central galaxy especially
in more massive haloes, boosting their size increase. Note also that faster mergers also imply less growth
time via star formation for satellites in intermediate to low mass haloes, thus proportionally decreasing
the sizes of the final remnants. This is the main reason
why galaxies in haloes with $\log \mhaloe/\msune \lesssim 13$ tend to be smaller than galaxies of similar
stellar mass in the model with longer \tdf\ (see also \figu\ref{fig|ReMstarRel}). The combination of these
two effects produces a steeper correlation of median sizes with respect to host halo masses (dotted lines)
than the same model with longer dynamical timescales (dot-dashed lines).

From \figu\ref{fig|ReMstarFixedMhaloNormSHAM} it also appears that the final sizes in the fast quenching model
(triple dot-dashed lines) are more compact with respect to a model with delayed quenching (dot-dashed lines),
and the effect is more relevant in lower mass host haloes. This result can be understood in our framework by recalling that
a faster quenching implies smaller and less massive satellites and a proportionally more contained growth for centrals.
Given that the specific star formation rate has a strong inverse dependence with stellar mass (\eq\ref{eq|SFR_Karim}),
lower mass galaxies grow proportionally more than more massive ones within the same interval of time.
Thus a faster quenching will more severely limit the growth of the lowest massive galaxies,
the latter being preferentially satellites in lower mass haloes.
We thus expect a faster satellite quenching to have, on average, a relatively more pronounced effect for
the growth of centrals in lower mass haloes.

\section{Discussion}
\label{sec|discu}

\begin{table*}
\begin{tabular}{|l|c|r|}
  \hline
  \hline
  \emph{PROCESS} & \emph{DESCRIPTION} & \emph{AFTER ERRORS} \\
  \hline
  \hline
  \textbf{DISC INSTABILITIES} & Mostly effective if violent and impulsive. & Only marginally reduced \\
  & Induce more compact
  bulges in less massive haloes  & \\
  & with lower circular velocities & \\
  \hline
  \textbf{MERGERS} & Only effective (at $z=0$) if short dynamical friction timescales. & Only marginally reduced \\
  & More effective (minor) galaxy mergers in more & \\
  & massive host haloes, thus larger centrals & \\
  \hline
  \textbf{GAS DISSIPATION} & Progressively more effective & Only marginally reduced \\
  & in less massive haloes with  & \\
  & gas-richer progenitors & \\
  \hline
  \textbf{SATELLITE EVOLUTION} & Overall milder effect. Present if & Significantly reduced \\
  & fast quenching/gas stripping, thus & \\
  & proportionally less growth in satellites in lower mass haloes. & \\
  & Induces more compact remnants in less massive haloes & \\
  \hline
  \end{tabular}
  \caption{List of main physical processes identified in this work which can cause environmental dependence
  in the median size of central, bulge-dominated early-type galaxies at fixed stellar mass. The second column briefly provides the main
  features characterizing each process within the context of environment. The third column details if the signal with environment, specifically induced by a given process, is still detectable after inclusion of systematic and statistical errors in size, stellar mass, and host halo mass, following the discussion in \sect\ref{subsec|comparedata}.}
  \label{table|models}
\end{table*}

\subsection{The physics behind environmental dependence}
\label{subsec|PhysicsOfEnvironment}

We have so far identified several physical effects which can cause a variation
in the median size of bulge-dominated central galaxies of similar stellar mass when moving from lower to higher mass host haloes.
Table\til\ref{table|models} contains a list of these processes, providing a brief description
for each one of them within the context of environment, and specifying if the signal with environment,
specifically induced by a given process, is still detectable after inclusion of systematic and statistical errors in size,
stellar mass, and host halo mass, following the discussion in \sect\ref{subsec|comparedata}.

We can summarize the physical processes identified in the previous sections as follows.
\begin{itemize}
  \item \textbf{Disc instabilities}. We discussed in reference to \figu\ref{fig|ReMstarFixedMhaloNorm}, that models
  characterized by strong and impulsive disc instabilities tend to grow massive bulges in less massive haloes.
  The instability criteria usually adopted in semi-analytic models in fact (e.g., \eq\ref{eq|discinstability}),
  are more easily met by massive galaxies in lower mass haloes, at fixed stellar mass. Being instabilities
  less efficient than mergers in building large bulges of same stellar mass (cfr. \eq\ref{eq|sizediscinstab}),
  this naturally increases the halo mass dependence in median sizes. Models characterized by strong disc
  instabilities can predict an increase in size of a factor of $\gtrsim 2$ when moving from field to clusters,
  difficult to reconcile with the data, even after convolution with substantial observational errors.
  We here stress, however, that violent disc instabilities, especially in high redshift, clumpy
  starforming discs could still play a substantial role in building stellar
  bulges \citep[e.g.,][]{Dekel09b,Bournaud11a,Bournaud11b}. If for example, violent
  disc instabilities are more common than previously thought \citep[e.g.,][]{Bournaud13,Mandelker13},
  than they can be triggered in different environments, thus reducing the tension with the data.
  What our findings seem to suggest
  is that the usually adopted analytic modelling for these types of processes (\sect\ref{sec|Models}),
  may still not be entirely appropriate for describing the complexities characterizing
  the different stages and modes of disc instabilities \citep[e.g.,][]{Atha13}.
  \item \textbf{Mergers/dynamical friction timescales}. We showed that even in the absence of strong disc instabilities,
      hierarchical models may still produce significant environmental dependence.
      This is particularly true for models with efficient merging, either induced by
      more bulge growth ({\sc morgana}), and/or lower dynamical friction timescales (B06, SHAM-low \tdf).
      For models with relatively longer \tdf\ (G11, SHAM), environmental dependence is minimal.
      This is mainly induced by the fact that, owing to the self-similarity of dark matter,
      the number of cumulative mergers down to $z=0$ on the central galaxy at fixed interval of stellar mass, does not largely
      increase when moving from low to high mass host haloes, with a moderate variation of up to a factor of $\lesssim 1.3-1.5$.
      This is clearly seen in
      \figu\ref{fig|Nmergers}, which plots the average total number of mergers for galaxies of
      stellar mass $11.5<\log \mstare/\msune<12$ as a function of host halo mass down to $z=0$.
      The left panel plots the expectations of our reference SHAM based on \tdf\ taken from
      the recent high-resolution simulations by \citealt{McCavana12},
      while the right ones of the SHAM run with the same dynamical friction timescales
      shortened by a factor of $1/3$.
      Central galaxies in more massive haloes have a tendency to have on average more minor mergers,
      while the number of major mergers, is roughly comparable in all environments \citep[see also, e.g.,][]{Hirschmann13}.
      This implies that centrals in more massive haloes
      will have a tendency for being larger. However, this size increase must also be relatively modest,
      as the cumulative number of mergers in the most massive haloes is at the most a factor of $\lesssim 1.3$ higher
      than in less massive haloes. On the other hand, the number of \emph{minor} (but not major) mergers increases by up to a
      factor of $\sim 4$ for the model characterized by lower \tdf. If the merger
      timescales are sufficiently short, then galaxies at the centres of clusters will tend to undergo more effective
      galaxy merging, thus naturally increasing their sizes and boosting any environmental dependence. Interestingly,
      lower dynamical friction timescales were favoured by, e.g., \citet{Newman12} to speed up the size growth of massive
      galaxies and improve the match with the data.
      We note that although merging may not be the dominant cause of environmental dependence at $z=0$, this does not exclude it
      may still induce a stronger environmental trend at higher redshifts. In fact, early-type galaxies born in denser
      environments are expected to undergo a boosted evolution at $z>1$,
      thus forming larger bulges at these epochs, in line with the observational evidence
      briefly summarized in \sect\ref{sec|intro} \citep[e.g.,][]{Delaye13}.
      We will further investigate the full evolution of bulge-dominated galaxies in different environments and their
      impact on size evolution in subsequent work (Shankar et al., in prep.).
      \begin{figure}
   \includegraphics[width=8.5truecm]{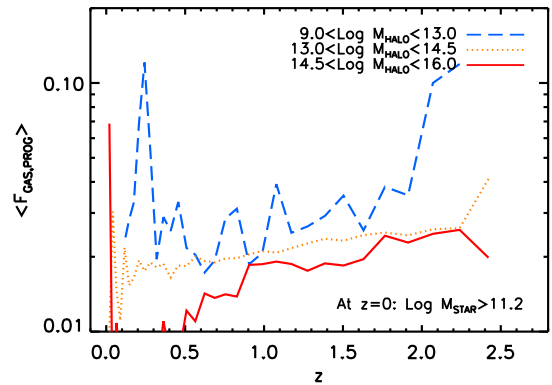}
    \caption{Predicted redshift evolution along the main branch of the median gas content in the progenitors before a major merger for galaxies with $\log \mstare/\msune >11.2$ at $z=0$, living at the centre of different host halo masses, as labelled. Massive galaxies in lower mass haloes tend to always be gas richer because the progenitors are less massive and thus gas richer.}
    \label{fig|Fgasevolution}
\end{figure}
  \item \textbf{Gas dissipation in (major) mergers}. Taken at face value, gas dissipation can produce a
  factor of $\sim 2$ increase in median size when moving from field to cluster environments,
  in apparent disagreement with the data (\figus\ref{fig|ReMstarFixedMhaloNorm} and \ref{fig|ReMstarFixedMhaloNormSHAM}).
      Within a given stellar mass bin, it is naturally expected that gas dissipation
      will more effectively shrink the sizes of the lowest mass and gas-richer galaxies (\eq\ref{eq|Redissipation}).
      This in turn will induce environmental dependence, given that lower mass galaxies preferentially live in lower mass haloes.
      However, even at fixed stellar mass galaxies residing in lower mass haloes will
      continue to have a higher probability to merge with massive, gas-richer satellites.
      This is because lower mass host haloes
      are more easily invested by lower stellar mass satellites with sufficiently high gas content to
      overcome the baryonic threshold for triggering a (dissipative) major merger (see \sect\ref{sec|Models}).
      To better visualize the latter effect,
      in \figu\ref{fig|Fgasevolution} we report the median total gas content of progenitors extracted from the SHAM model.
      We consider the gas fractions in the pre-merger phase as a function of host halo mass, weighted in a way to
      minimize any additional spurious trend due to the stellar mass dependence in the input gas fractions (\eq\ref{eq|fgasStewart}).
      We first compute the quantity
      \begin{equation}
      F_{\rm gas,merg}=M_{\rm gas,prog}/(M_{\rm star,prog}+M_{\rm star,burst})\, ,
        \label{eq|FgasMerg}
      \end{equation}
      which is the ratio between total gas content of progenitors in major mergers, and the sum of the total stellar mass
      of the progenitors plus the amount of mass formed during the merger (computed via \eq\ref{eq|eburst}).
      The latter quantity is just the fraction used in \eq\ref{eq|Redissipation}.
      The higher the gas fraction $F_{\rm gas,merg}$, the more compact the remnant will be.
      To then make a proper comparison among progenitors of different mass, similarly to what we do for sizes
      we weight $F_{\rm gas,merg}$ with respect to the remnant's stellar mass as
       \begin{equation}
      \log F_{\rm gas,prog}=\log F_{\rm gas,merg}+\alpha(z)[11-\log \mstare]\, ,
        \label{eq|FgasProg}
      \end{equation}
      with $\alpha(z)$ the same as in \eq\ref{eq|fgasStewart}.
      The result is reported in \figu\ref{fig|Fgasevolution}, which shows that the progenitors of centrals in less massive haloes are
      always gas richer. This in turn explains why the remnants are proportionally more compact,
      as gas dissipation produces more compact remnants, the gas-richer the merging progenitors (\eq\ref{eq|Redissipation}).
      The non ideal performance of models with gas dissipation is particularly intriguing. The latter
      process has been recognized to play a significant role in hydro-simulations
      \citep{Hop08FP,Covington11}, and also to help in better reproducing other scaling relations, such as the ones with age
      and velocity dispersion \citep{Shankar13}, or possibly also between bulge size and stellar mass, as discussed in
      \sect\ref{subsec|SizeMstarRelation}.
      On the other hand, as mentioned in \sect\ref{subsec|SizeMstarRelation},
      a properly fine-tuned disc regrowth/survival mechanism
      may suitably increase the total sizes of remnants in gas-richer environments, thus
      helping to better match the \re-\mstar\ relation (\figu\ref{fig|ReMstarRel}),
      and at the same time reduce environmental dependence. We will explore such possibilities in future work.
  \item \textbf{Evolution of satellites: gas consumption, quenching, and stripping}.
      Including quenching in the model implies restricting the growth of infalling satellites.
      In turn, fast quenching tends to produce more compact
      remnants at fixed stellar mass with respect to the data, although the difference could still
      be within the 1\til$\sigma$ uncertainty (cfr. \figu\ref{fig|ReMstarSHAMs}).
      More interestingly, we pointed out with respect to \figu\ref{fig|ReMstarFixedMhaloNormSHAM},
      that a model with fast quenching
      predicts some enhanced environmental size dependence with respect to a model with slower quenching.
      As shown in \sect\ref{subsec|varyingSHAMs}, the latter effect can be broadly understood along the following lines.
      In general, galaxies of a given stellar mass residing in less massive haloes,
      will preferentially merge with lower mass, gas-richer, and more star-forming satellites,
      the latter two features being a direct consequence of the anti-correlation between gas fractions and
      specific star formation rate with stellar mass
      (Eqs. \ref{eq|fgasStewart} and \ref{eq|SFR_Karim}, respectively).
      In other words, within the same amount of time, lower mass satellites grow proportionally
      more than more massive ones.
      In a fast quenching scenario, however, satellites' growth is inhibited, proportionally more in lower mass haloes
      invested by the least massive satellites. Centrals in lower mass haloes are thus expected to grow less in size
      with respect to their counterparts in more massive haloes.
      Indeed, we have verified that the number of (minor) mergers (down to 1\% in progenitors' mass ratio),
      is reduced by $\sim (30 \pm 20)\%$ for centrals in haloes $\log \mhaloe/\msune \lesssim 13$
      with respect to the reference model in the left panel of \figu\ref{fig|Nmergers}, thus proportionally preventing
      their size growth.
      The fast quenching process, however, tends to be weaker than the previous ones in the list,
      since the difference in number of mergers between less and more massive haloes is less than $1.5$ $\sigma$.
      The net effect is nevertheless visible in the top left panel of \figu\ref{fig|ReMstarFixedMhaloNormSHAM}, where
      central galaxies in haloes below $\log \mhaloe/\msune \lesssim 13$ appear
      progressively more compact in the fast quenching model (triple dot-dashed line) than in the slower quenching one (solid line).
      Finally, as discussed in Appendix A,
      a model with faster quenching will tend to retain larger fractions of gas with respect to what actually
      observed, at least at this basic level of the modelling.
\end{itemize}

\subsection{Scatter around the $R_e-\mstare$ relation}
\label{subsec|ScatterReMstarAndBTfractions}

\begin{figure*}
    \includegraphics[width=15truecm]{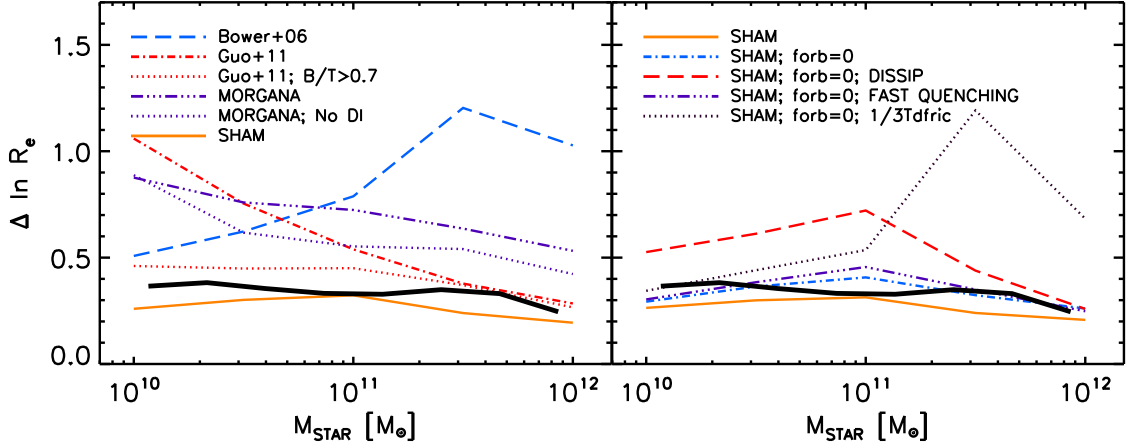}
    \caption{Predicted scatter around the median size-stellar mass relation
    for different models (\emph{left}), and for different variations within the SHAM (\emph{right}).
    Data are extracted from the \citet{Huertas13b} catalogue.
    Most models predict a large scatter. However, a large fraction of this scatter is caused
    by galaxies with lower $B/T$. Bulge-dominated galaxies,
    with $B/T>0.7$, tend to better line-up with the data. SHAMs, characterized by having
    tighter relations in the scaling relations of progenitors, tend to provide a better match to the data (see text for details).}
    \label{fig|RezScatters}
\end{figure*}

So far we mainly focused on the slopes of the size-stellar mass relations
and discussed dispersions around this relation only in terms of dependence on host halo mass.
We here discuss model predictions in terms of \emph{total} scatter around the median.
We should immediately emphasize that studying total scatter around the
median size-stellar mass relation, i.e., its global 1\til$\sigma$ dispersion,
although possibly correlated, is not necessarily
equivalent to study environmental dependence as we did in the previous sections. In fact,
until now our concern was focused on probing departures from the median
of specific subclasses of galaxies, labelled by different host halo masses.
Such deviations were further weighted to take into account
the exact location of each galaxy on the relation,
irrespective of how large the bin in stellar mass considered. This methodology
can thus easily produce dispersions different from the canonical 1\til$\sigma$, rigourously
computed for all galaxies in a narrow bin of stellar mass. For example, \citet{Shankar13} showed that the
G11 model with or without dissipation produces very similar global dispersions around the mean relation, especially
for the most massive galaxies (their Fig. 5),
despite the inclusion of dissipation
inducing an environmental dependence systematically higher by a factor of $\gtrsim 1.5$ (\figu\ref{fig|ReMstarFixedMhaloNorm}).
As an additional proof, we have also checked that all the results on the total scatter presented below
do not change significantly if, for example, we exclude from the analysis
massive groups and clusters. Limiting the sample to haloes less massive than, say,
$\mhaloe \lesssim 3\times 10^{13}\, \msune$, yields in fact nearly identical results, except
for entirely cutting out galaxies in the highest bins of stellar mass, as expected. We will thus proceed discussing the results
on global scatter mostly as a separate issue with respect to environmental dependence,
although we will highlight connections where relevant.

\figu\ref{fig|RezScatters} compares the outcomes of our reference hierarchical
galaxy models with respect to the data (solid/black line).
We confirm previous claims \citep[e.g.,][]{Nair11,Bernardi13}
for an extremely tight correlation in the observed relation with $\Delta \ln R_e \sim 0.35$, i.e., just
$\sim 0.15$ dex. \citet{Bernardi13} have recently showed via Monte Carlo simulations that the intrinsic, true scatter should
in fact be even smaller by a factor $\gtrsim 1.5-2$ (cfr. their Fig. 13).

All hierarchical models instead tend to predict much larger dispersions than those observed.
Here we only focus on the raw predictions of the models but,
as anticipated in \sect\ref{subsec|comparedata}, any convolution with correlated and statistical errors
would enhance the predicted scatter by up to $\sim 30\%$, clearly worsening the comparison to the data.
The causes behind such broadening are multiple. First, disc instabilities can effectively
increase the scatter in the scaling relations involving sizes. If modelled as in \eq\ref{eq|sizediscinstab},
disc instabilities will always be less efficient than mergers in building large bulges,
irrespective of the type of disc instability considered, either moderate ones as in G11, or stronger ones as in B06.
Thus, similarly to what discussed in reference to environmental dependence,
at fixed stellar mass disc instabilities will also generate a larger fraction of
more compact bulges, increasing the scatter.
The left panel of \figu\ref{fig|RezScatters} shows that the large size distribution in the G11 model (dot-dashed, red line)
for stellar masses below $\mstare\lesssim 10^{11}\, \msune$,
is mainly caused by stellar bar instabilities. In line with what
highlighted by \citet{Shankar13}, a large fraction of bulges in this mass range is built via secular processes.
Restricting instead to bulge-dominated galaxies with \bt$>0.7$ (red, dotted line), with mostly merger-driven growth
\citep{Shankar13}, neatly cuts out all outliers beyond $\Delta \ln R_e \gtrsim 0.5$. Similarly, the {\sc morgana} model
(triple dot-dashed, purple line)
also predicts a large scatter, which is significantly reduced to $\Delta \ln R_e \lesssim 0.6$ for massive galaxies
if disc instabilities are not included (dotted, purple line).
The B06 model predicts large dispersions with respect to the data, at all stellar masses.
We believe that at least part of this discrepancy is due to the strong disc instabilities and the
relatively lower dynamical friction timescales
included in this model (see \sect\ref{subsec|B06model}).

Even after removing disc instabilities, models still tend to predict larger
scatters than observed. This is clear from the left panel of \figu\ref{fig|RezScatters} as
both the {\sc morgana} and the G11 models (purple and red dotted lines, respectively)
still lie above the data at all stellar masses.
It was already pointed out that the chaotic nature of galaxy merger trees could in fact generate too large
size distributions at fixed stellar mass with respect to what actually measured in
the local Universe \citep[e.g.,][]{Nipoti09,Nair11,Shankar13}.
However, the disagreement with the data although significant, is not large.
Indeed, we find that the SHAM, despite being built in a very similar hierarchical context as for the other models,
predicts an exceptionally limited dispersion, actually somewhat lower than the observed one,
and possibly comparable to the intrinsic one claimed by \citet{Bernardi13}.
We checked in fact that, at least above $\log \mstare/\msune \gtrsim 11.2$,
the predicted scatter from the SHAM inclusive of all errors amounts to $\Delta \ln R_e \sim 0.3$,
in close agreement to the observed one.
The success of the latter model relies on the very tight input scaling
relations in mass and size defining the progenitor disc galaxies.
For example, as detailed in \sect\ref{subsec|SHAM},
in our reference model we have assumed discs to follow an empirical, time-dependent disc size-stellar mass relation, with an intrinsic
Gaussian scatter of 0.1 dex. We checked that increasing this scatter, proportionally boosts the
final dispersion around the size-stellar mass relation of the remnant bulge-dominated galaxies.
In other words, galaxy merger trees may not necessarily broaden the input correlations.
A similar conclusion has been recently reached by \citet{Taranu13},
who demonstrated via collisionless simulations of dry mergers in group environments,
that stochastic merging can indeed produce tight scaling relations for
early-type remnants, as long as the merging galaxies also follow tight scaling relations.

We find the latter conclusion to be quite robust against variations in the input assumptions
of the SHAM model. By varying orbital energies (\forb), or the time for quenching, only mildly
impacts the resulting scatter in sizes (dot-dashed, blue and triple dot-dashed, purple lines, respectively).
The only notable exceptions are variations in \tdf\ and, possibly, gas dissipation.
Despite the short \tdf\ model (dotted, black line, right panel) being identical to the reference SHAM (i.e., very
tight scaling relations for the progenitors), the resulting correlations for bulges appear quite broader,
especially at masses above $\gtrsim 10^{11}\, \msune$, which are the ones most affected by mergers.
Indeed, when mergers become too numerous they can induce chaotic behaviours in the
resulting scaling relations, in line with previous claims. However, when proper (sufficiently long) \tdf\ are
adopted, alongside with tight scaling relations for progenitors, then mergers can preserve
tight scaling relations for remnants.

The SHAM model with gas dissipation (red, long-dashed line, right panel)
tends to produce a scatter larger than the
ones without dissipation around $\mstare \lesssim 2\times 10^{11}\, \msune$, but comparable
at increasingly higher stellar masses, broadly in line with
what claimed by \citet{Shankar13}. We can partly ascribe the increase in dispersion
at intermediate and lower stellar masses to the slow quenching timescales assumed
in the reference SHAM. We checked in fact that by increasing the quenching and/or the stripping
would better limit the growth of the scatter in sizes, especially in the model with dissipation.

Overall, what is most relevant to the present discussion
is that mergers alone do not necessarily imply larger scatters in size distributions
with respect to what observed. As long as the dynamical timescales are sufficiently long,
comparable to what suggested by detailed $N$-body simulations, and the scaling relations
of progenitors are sufficiently tight, the complex mixture of mergers may still preserve a contained scatter in the remnants,
closer to what is observed.

\section{Conclusions}
\label{sec|conclu}

In this work we have compared state-of-the-art semi-analytic models
of galaxy formation as well as advanced sub-halo abundance matching (SHAM) models,
with a large sample of early-type galaxies from SDSS.
In particular, we focused our attention on the
environmental dependence of sizes of central galaxies as a function of host halo mass.
In the data, information on host halo mass is derived by cross-correlating
the SDSS morphological sample by \citet{Huertas13a} with the \citet{Yang07} halo catalog.
Sizes are derived from S\'{e}rsic fits to the SDSS images \citep{Bernardi13}.
We then selected early-type galaxies with $\log \mstare/\msune > 11.2$.
We find a flat distribution of median size as a function of host halo mass,
in line with previous studies in the local Universe \citep[e.g.,][]{Guo09,Huertas13b}.

All hierarchical models considered in this work instead tend to predict a moderate to strong
environmental dependence, with the median size difference of a factor of $\sim 1.5-3$ when
moving from the lowest ($\gtrsim 3 \times 10^{12}\, \msune$) to the highest ($\sim 10^{15}\, \msune$) host halo masses.
At face value the discrepancy
with the data is highly significant. However, the convolution with the
(correlated) errors in the observations, can wash out part of the trends with host halo mass
predicted by some models, thus lowering the significance of the discrepancy.
We however find that those models which predict a difference higher than a factor of $\gtrsim 1.5-2$,
tend to preserve the signal in samples with the same number of galaxies as in SDSS.

Despite the observational uncertainties, the data tend to disfavour hierarchical models
characterized by strong and impulsive disc instabilities, strong gas dissipation in major mergers,
short dynamical friction timescales, and very short quenching timescales in infalling satellites.
These results hold irrespective of the model adopted, semi-analytic or semi-empirical.
Interestingly, mergers at the rate predicted by N-body simulations are not a major cause
for environmental dependence in the local Universe,
because the cumulative number of mergers on the central galaxies down to $z=0$,
and thus their related size growth, is not a strong function of host halo mass at fixed bin of stellar mass.

Galaxies residing in less massive haloes are preferentially involved in mergers
with gas richer satellites, thus inducing proportionally more
gas dissipation, more compact remnants and more environmental dependence.
Also, galaxies residing in less massive haloes more frequently meet the condition for disc
instabilities (massive discs in relatively less massive haloes), triggering the growth of bulges
with smaller sizes with respect to equally massive bulges grown via mergers, thus further increasing
any environmental dependence.
Finally, if the quenching of satellites is sufficiently rapid, then this will impact
more efficiently less massive galaxies with the highest specific star formation rates.
In turn, central galaxies residing in less massive haloes,
involved preferentially in minor mergers with the least massive satellites, will grow proportionally
less in size than their counterparts in more massive haloes, thus inducing additional environmental dependence.
We discuss possibilities to alleviate tensions between models and the data in \sect\ref{sec|discu}.

We also discussed additional key issues related to sizes and environment in bulge dominated galaxies
and to the hierarchical models considered above.
Most importantly, we showed that the size-stellar mass relation of local galaxies can be well reproduced by hierarchical models
both in slope and scatter as long as the scaling relations of the disc progenitors are sufficiently tight,
and the dynamical friction timescales are sufficiently long.

It will be interesting to discuss detailed model predictions to higher redshifts where,
as referenced in \sect\ref{sec|intro},
there is now growing evidence for an accelerated structural evolution of massive
galaxies in denser environments.
In fact merging could still induce a faster evolution of galaxies
in very dense regions, which could create an apparent trend with environment at high redshifts.
We leave the exploration of these issues to separate work (Shankar et al., in prep.).

Environment continues to play a significant role in constraining galaxy evolution. When future dynamical and spectral
observations of massive, early-type galaxies and their surroundings will be able to further tighten
the measurements on size, stellar mass, and halo mass, the constraints
will become invaluable to discern among the successful galaxy formation models.

\section*{Acknowledgments}
FS warmly thanks Simona Gattorno for all her kind and continuous support.
FS acknowledges support from a Marie Curie Grant.
FF acknowledges financial support from the Klaus Tschira Foundation
and the Deutsche Forschungsgemeinschaft through Transregio 33, ``The
Dark Universe''. We thank an anonymous referee for a constructive report
that helped improving the presentation of the results.

\bibliographystyle{mn2e_Daly}
\bibliography{../../../RefMajor_Rossella}

\begin{thebibliography}{123}
\expandafter\ifx\csname natexlab\endcsname\relax\def\natexlab#1{#1}\fi

\bibitem[{{Abazajian} {et~al}\mbox{.}(2009){Abazajian}, {Adelman-McCarthy},
  {Ag{\"u}eros}, {Allam}, {Allende Prieto}, {An}, {Anderson}, {Anderson},
  {Annis}, {Bahcall}, \& et~al.}]{2009ApJS..182..543A}
{Abazajian} K.~N. {et~al.}, 2009, \apjs, 182, 543

\bibitem[{{Aguerri} {et~al}\mbox{.}(2007){Aguerri}, {S{\'a}nchez-Janssen}, \&
  {Mu{\~n}oz-Tu{\~n}{\'o}n}}]{Aguerri07}
{Aguerri} J.~A.~L., {S{\'a}nchez-Janssen} R., {Mu{\~n}oz-Tu{\~n}{\'o}n} C.,
  2007, \aap, 471, 17

\bibitem[{{Angulo} {et~al}\mbox{.}(2009){Angulo}, {Lacey}, {Baugh}, \&
  {Frenk}}]{Angulo09}
{Angulo} R.~E., {Lacey} C.~G., {Baugh} C.~M., {Frenk} C.~S., 2009, \mnras, 399,
  983

\bibitem[{{Athanassoula} {et~al}\mbox{.}(2013){Athanassoula}, {Machado}, \&
  {Rodionov}}]{Atha13}
{Athanassoula} E., {Machado} R.~E.~G., {Rodionov} S.~A., 2013, \mnras, 429,
  1949

\bibitem[{{Bakos} \& {Trujillo}(2012)}]{BakosTrujillo}
{Bakos} J., {Trujillo} I., 2012, arXiv:1204.3082

\bibitem[{{Bassett} {et~al}\mbox{.}(2013){Bassett}, {Papovich}, {Lotz}, {Bell},
  {Finkelstein}, {Newman}, {Tran}, {Almaini}, {Lani}, {Cooper}, {Croton},
  {Dekel}, {Ferguson}, {Kocevski}, {Koekemoer}, {Koo}, {McGrath}, {McIntosh},
  \& {Wechsler}}]{Bassett13}
{Bassett} R. {et~al.}, 2013, \apj, 770, 58

\bibitem[{{Behroozi} {et~al}\mbox{.}(2010){Behroozi}, {Conroy}, \&
  {Wechsler}}]{Behroozi10}
{Behroozi} P.~S., {Conroy} C., {Wechsler} R.~H., 2010, \apj, 717, 379

\bibitem[{{Bernardi}(2009)}]{Bernardi09}
{Bernardi} M., 2009, \mnras, 395, 1491

\bibitem[{{Bernardi} {et~al}\mbox{.}(2013){Bernardi}, {Meert}, {Sheth},
  {Vikram}, {Huertas-Company}, {Mei}, \& {Shankar}}]{Bernardi13}
{Bernardi} M., {Meert} A., {Sheth} R.~K., {Vikram} V., {Huertas-Company} M.,
  {Mei} S., {Shankar} F., 2013, \mnras, 436, 697

\bibitem[{{Bernardi} {et~al}\mbox{.}(2012){Bernardi}, {Meert}, {Vikram},
  {Huertas-Company}, {Mei}, {Shankar}, \& {Sheth}}]{Bernardi12}
{Bernardi} M., {Meert} A., {Vikram} V., {Huertas-Company} M., {Mei} S.,
  {Shankar} F., {Sheth} R.~K., 2012, arXiv:1211.6122

\bibitem[{{Bernardi} {et~al}\mbox{.}(2011{\natexlab{a}}){Bernardi}, {Roche},
  {Shankar}, \& {Sheth}}]{Bernardi11a}
{Bernardi} M., {Roche} N., {Shankar} F., {Sheth} R.~K., 2011{\natexlab{a}},
  \mnras, 412, 684

\bibitem[{{Bernardi} {et~al}\mbox{.}(2011{\natexlab{b}}){Bernardi}, {Roche},
  {Shankar}, \& {Sheth}}]{Bernardi11b}
{Bernardi} M., {Roche} N., {Shankar} F., {Sheth} R.~K., 2011{\natexlab{b}},
  \mnras, 412, L6

\bibitem[{{Bernardi} {et~al}\mbox{.}(2010){Bernardi}, {Shankar}, {Hyde}, {Mei},
  {Marulli}, \& {Sheth}}]{Bernardi10}
{Bernardi} M., {Shankar} F., {Hyde} J.~B., {Mei} S., {Marulli} F., {Sheth}
  R.~K., 2010, \mnras, 404, 2087

\bibitem[{{Bernardi} {et~al}\mbox{.}(2003){Bernardi}, {Sheth}, {Annis},
  {Burles}, {Eisenstein}, {Finkbeiner}, {Hogg}, {Lupton}, {Schlegel},
  {SubbaRao}, {Bahcall}, {Blakeslee}, {Brinkmann}, {Castander}, {Connolly},
  {Csabai}, {Doi}, {Fukugita}, {Frieman}, {Heckman}, {Hennessy}, {Ivezi{\'c}},
  {Knapp}, {Lamb}, {McKay}, {Munn}, {Nichol}, {Okamura}, {Schneider}, {Thakar},
  \& {York}}]{Bernardi03I}
{Bernardi} M. {et~al.}, 2003, \aj, 125, 1817

\bibitem[{{Bournaud} {et~al}\mbox{.}(2011{\natexlab{a}}){Bournaud}, {Chapon},
  {Teyssier}, {Powell}, {Elmegreen}, {Elmegreen}, {Duc}, {Contini}, {Epinat},
  \& {Shapiro}}]{Bournaud11a}
{Bournaud} F. {et~al.}, 2011{\natexlab{a}}, \apj, 730, 4

\bibitem[{{Bournaud} {et~al}\mbox{.}(2011{\natexlab{b}}){Bournaud}, {Dekel},
  {Teyssier}, {Cacciato}, {Daddi}, {Juneau}, \& {Shankar}}]{Bournaud11b}
{Bournaud} F., {Dekel} A., {Teyssier} R., {Cacciato} M., {Daddi} E., {Juneau}
  S., {Shankar} F., 2011{\natexlab{b}}, \apjl, 741, L33

\bibitem[{{Bournaud} {et~al}\mbox{.}(2013){Bournaud}, {Perret}, {Renaud},
  {Dekel}, {Elmegreen}, {Elmegreen}, {Teyssier}, {Amram}, {Daddi}, {Duc},
  {Elbaz}, {Epinat}, {Gabor}, {Juneau}, {Kraljic}, \& {Le Floch'}}]{Bournaud13}
{Bournaud} F. {et~al.}, 2013, arXiv:1307.7136

\bibitem[{{Bower} {et~al}\mbox{.}(2006){Bower}, {Benson}, {Malbon}, {Helly},
  {Frenk}, {Baugh}, {Cole}, \& {Lacey}}]{Bower06}
{Bower} R.~G., {Benson} A.~J., {Malbon} R., {Helly} J.~C., {Frenk} C.~S.,
  {Baugh} C.~M., {Cole} S., {Lacey} C.~G., 2006, \mnras, 370, 645

\bibitem[{{Boylan-Kolchin} {et~al}\mbox{.}(2008){Boylan-Kolchin}, {Ma}, \&
  {Quataert}}]{Boylan08}
{Boylan-Kolchin} M., {Ma} C.-P., {Quataert} E., 2008, \mnras, 383, 93

\bibitem[{{Bruzual} \& {Charlot}(2003)}]{BC03}
{Bruzual} G., {Charlot} S., 2003, \mnras, 344, 1000

\bibitem[{{Carollo} {et~al}\mbox{.}(2013){Carollo}, {Bschorr}, {Renzini},
  {Lilly}, {Capak}, {Cibinel}, {Ilbert}, {Onodera}, {Scoville}, {Cameron},
  {Mobasher}, {Sanders}, \& {Taniguchi}}]{Carollo13}
{Carollo} C.~M. {et~al.}, 2013, \apj, 773, 112

\bibitem[{{Catinella} {et~al}\mbox{.}(2010){Catinella}, {Schiminovich},
  {Kauffmann}, {Fabello}, {Wang}, {Hummels}, {Lemonias}, {Moran}, {Wu},
  {Giovanelli}, {Haynes}, {Heckman}, {Basu-Zych}, {Blanton}, {Brinchmann},
  {Budav{\'a}ri}, {Gon{\c c}alves}, {Johnson}, {Kennicutt}, {Madore}, {Martin},
  {Rich}, {Tacconi}, {Thilker}, {Wild}, \& {Wyder}}]{Catinella10}
{Catinella} B. {et~al.}, 2010, \mnras, 403, 683

\bibitem[{{Cattaneo} {et~al}\mbox{.}(2011){Cattaneo}, {Mamon}, {Warnick}, \&
  {Knebe}}]{Cattaneo11}
{Cattaneo} A., {Mamon} G.~A., {Warnick} K., {Knebe} A., 2011, \aap, 533, A5

\bibitem[{{Chabrier}(2003)}]{Chabrier03}
{Chabrier} G., 2003, \pasp, 115, 763

\bibitem[{{Chiosi} {et~al}\mbox{.}(2012){Chiosi}, {Merlin}, \&
  {Piovan}}]{Chiosi12}
{Chiosi} C., {Merlin} E., {Piovan} L., 2012, arXiv:1206.2532

\bibitem[{{Cimatti} {et~al}\mbox{.}(2012){Cimatti}, {Nipoti}, \&
  {Cassata}}]{Cimatti12}
{Cimatti} A., {Nipoti} C., {Cassata} P., 2012, \mnras, 422, L62

\bibitem[{{Cole} {et~al}\mbox{.}(2000){Cole}, {Lacey}, {Baugh}, \&
  {Frenk}}]{Cole00}
{Cole} S., {Lacey} C.~G., {Baugh} C.~M., {Frenk} C.~S., 2000, \mnras, 319, 168

\bibitem[{{Cooper} {et~al}\mbox{.}(2012){Cooper}, {Griffith}, {Newman}, {Coil},
  {Davis}, {Dutton}, {Faber}, {Guhathakurta}, {Koo}, {Lotz}, {Weiner},
  {Willmer}, \& {Yan}}]{Cooper12}
{Cooper} M.~C. {et~al.}, 2012, \mnras, 419, 3018

\bibitem[{{Covington} {et~al}\mbox{.}(2011){Covington}, {Primack}, {Porter},
  {Croton}, {Somerville}, \& {Dekel}}]{Covington11}
{Covington} M.~D., {Primack} J.~R., {Porter} L.~A., {Croton} D.~J.,
  {Somerville} R.~S., {Dekel} A., 2011, \mnras, 415, 3135

\bibitem[{{De Lucia} \& {Blaizot}(2007)}]{DeLuciaBlaizot}
{De Lucia} G., {Blaizot} J., 2007, \mnras, 375, 2

\bibitem[{{De Lucia} {et~al}\mbox{.}(2010){De Lucia}, {Boylan-Kolchin},
  {Benson}, {Fontanot}, \& {Monaco}}]{DeLucia10}
{De Lucia} G., {Boylan-Kolchin} M., {Benson} A.~J., {Fontanot} F., {Monaco} P.,
  2010, \mnras, 406, 1533

\bibitem[{{De Lucia} {et~al}\mbox{.}(2011){De Lucia}, {Fontanot}, {Wilman}, \&
  {Monaco}}]{DeLucia11}
{De Lucia} G., {Fontanot} F., {Wilman} D., {Monaco} P., 2011, \mnras, 517

\bibitem[{{De Lucia} {et~al}\mbox{.}(2006){De Lucia}, {Springel}, {White},
  {Croton}, \& {Kauffmann}}]{DeLucia06}
{De Lucia} G., {Springel} V., {White} S.~D.~M., {Croton} D., {Kauffmann} G.,
  2006, \mnras, 366, 499

\bibitem[{{de Vaucouleurs}(1948)}]{deVac}
{de Vaucouleurs} G., 1948, Annales d'Astrophysique, 11, 247

\bibitem[{{Dekel} {et~al}\mbox{.}(2009){Dekel}, {Sari}, \&
  {Ceverino}}]{Dekel09b}
{Dekel} A., {Sari} R., {Ceverino} D., 2009, \apj, 703, 785

\bibitem[{{Delaye} {et~al}\mbox{.}(2013){Delaye}, {Huertas-Company}, {Mei},
  {Lidman}, {Licitra}, {Newman}, {Raichoor}, {Shankar}, {Barrientos},
  {Bernardi}, {Cerulo}, {Couch}, {Demarco}, {Mu{\~n}oz}, {Sanchez-Janssen}, \&
  {Tanaka}}]{Delaye13}
{Delaye} L. {et~al.}, 2013, arXiv:1307.0003

\bibitem[{{Efstathiou} {et~al}\mbox{.}(1982){Efstathiou}, {Lake}, \&
  {Negroponte}}]{Efstathiou82}
{Efstathiou} G., {Lake} G., {Negroponte} J., 1982, \mnras, 199, 1069

\bibitem[{{Fan} {et~al}\mbox{.}(2010){Fan}, {Lapi}, {Bressan}, {Bernardi}, {De
  Zotti}, \& {Danese}}]{Fan10}
{Fan} L., {Lapi} A., {Bressan} A., {Bernardi} M., {De Zotti} G., {Danese} L.,
  2010, \apj, 718, 1460

\bibitem[{{Fontanot} {et~al}\mbox{.}(2011){Fontanot}, {De Lucia}, {Wilman}, \&
  {Monaco}}]{Fontanot11}
{Fontanot} F., {De Lucia} G., {Wilman} D., {Monaco} P., 2011, \mnras, 416, 409

\bibitem[{{Fu} {et~al}\mbox{.}(2010){Fu}, {Guo}, {Kauffmann}, \&
  {Krumholz}}]{Fu10}
{Fu} J., {Guo} Q., {Kauffmann} G., {Krumholz} M.~R., 2010, \mnras, 409, 515

\bibitem[{{Fu} {et~al}\mbox{.}(2013){Fu}, {Kauffmann}, {Huang}, {Yates},
  {Moran}, {Heckman}, {Dav{\'e}}, {Guo}, \& {Henriques}}]{Fu13}
{Fu} J. {et~al.}, 2013, \mnras

\bibitem[{{Gonz{\'a}lez} {et~al}\mbox{.}(2009){Gonz{\'a}lez}, {Lacey}, {Baugh},
  {Frenk}, \& {Benson}}]{Gonzalez09}
{Gonz{\'a}lez} J.~E., {Lacey} C.~G., {Baugh} C.~M., {Frenk} C.~S., {Benson}
  A.~J., 2009, \mnras, 397, 1254

\bibitem[{{Guo} {et~al}\mbox{.}(2013){Guo}, {White}, {Angulo}, {Henriques},
  {Lemson}, {Boylan-Kolchin}, {Thomas}, \& {Short}}]{Guo13}
{Guo} Q., {White} S., {Angulo} R.~E., {Henriques} B., {Lemson} G.,
  {Boylan-Kolchin} M., {Thomas} P., {Short} C., 2013, \mnras, 428, 1351

\bibitem[{{Guo} {et~al}\mbox{.}(2011){Guo}, {White}, {Boylan-Kolchin}, {De
  Lucia}, {Kauffmann}, {Lemson}, {Li}, {Springel}, \& {Weinmann}}]{Guo11}
{Guo} Q. {et~al.}, 2011, \mnras, 164

\bibitem[{{Guo} {et~al}\mbox{.}(2010){Guo}, {White}, {Li}, \&
  {Boylan-Kolchin}}]{Guo10}
{Guo} Q., {White} S., {Li} C., {Boylan-Kolchin} M., 2010, \mnras, 404, 1111

\bibitem[{{Guo} {et~al}\mbox{.}(2009){Guo}, {McIntosh}, {Mo}, {Katz}, {van den
  Bosch}, {Weinberg}, {Weinmann}, {Pasquali}, \& {Yang}}]{Guo09}
{Guo} Y. {et~al.}, 2009, \mnras, 398, 1129

\bibitem[{{Harker} {et~al}\mbox{.}(2006){Harker}, {Cole}, {Helly}, {Frenk}, \&
  {Jenkins}}]{Harker06}
{Harker} G., {Cole} S., {Helly} J., {Frenk} C., {Jenkins} A., 2006, \mnras,
  367, 1039

\bibitem[{{Henriques} {et~al}\mbox{.}(2012){Henriques}, {White}, {Lemson},
  {Thomas}, {Guo}, {Marleau}, \& {Overzier}}]{Henriques12}
{Henriques} B.~M.~B., {White} S.~D.~M., {Lemson} G., {Thomas} P.~A., {Guo} Q.,
  {Marleau} G.-D., {Overzier} R.~A., 2012, \mnras, 421, 2904

\bibitem[{{Hirschmann} {et~al}\mbox{.}(2013){Hirschmann}, {De Lucia}, {Iovino},
  \& {Cucciati}}]{Hirschmann13}
{Hirschmann} M., {De Lucia} G., {Iovino} A., {Cucciati} O., 2013, \mnras, 433,
  1479

\bibitem[{{Hopkins} {et~al}\mbox{.}(2010){Hopkins}, {Croton}, {Bundy},
  {Khochfar}, {van den Bosch}, {Somerville}, {Wetzel}, {Keres}, {Hernquist},
  {Stewart}, {Younger}, {Genel}, \& {Ma}}]{HopkinsMergers}
{Hopkins} P.~F. {et~al.}, 2010, \apj, 724, 915

\bibitem[{{Hopkins} {et~al}\mbox{.}(2009){Hopkins}, {Hernquist}, {Cox},
  {Keres}, \& {Wuyts}}]{Hop08FP}
{Hopkins} P.~F., {Hernquist} L., {Cox} T.~J., {Keres} D., {Wuyts} S., 2009,
  \apj, 691, 1424

\bibitem[{{Huertas-Company} {et~al}\mbox{.}(2011){Huertas-Company}, {Aguerri},
  {Bernardi}, {Mei}, \& {S{\'a}nchez Almeida}}]{Huertas11}
{Huertas-Company} M., {Aguerri} J.~A.~L., {Bernardi} M., {Mei} S., {S{\'a}nchez
  Almeida} J., 2011, \aap, 525, A157

\bibitem[{{Huertas-Company}
  {et~al}\mbox{.}(2013{\natexlab{a}}){Huertas-Company}, {Mei}, {Shankar},
  {Delaye}, {Raichoor}, {Covone}, {Finoguenov}, {Kneib}, {Le}, \&
  {Povic}}]{Huertas13a}
{Huertas-Company} M. {et~al.}, 2013{\natexlab{a}}, \mnras, 428, 1715

\bibitem[{{Huertas-Company}
  {et~al}\mbox{.}(2013{\natexlab{b}}){Huertas-Company}, {Shankar}, {Mei},
  {Bernardi}, {Aguerri}, {Meert}, \& {Vikram}}]{Huertas13b}
{Huertas-Company} M., {Shankar} F., {Mei} S., {Bernardi} M., {Aguerri}
  J.~A.~L., {Meert} A., {Vikram} V., 2013{\natexlab{b}}, \apj, 779, 29

\bibitem[{{Ishibashi} {et~al}\mbox{.}(2013){Ishibashi}, {Fabian}, \&
  {Canning}}]{Ishi13}
{Ishibashi} W., {Fabian} A.~C., {Canning} R.~E.~A., 2013, \mnras, 431, 2350

\bibitem[{{Karim} {et~al}\mbox{.}(2011){Karim}, {Schinnerer},
  {Mart{\'{\i}}nez-Sansigre}, {Sargent}, {van der Wel}, {Rix}, {Ilbert},
  {Smol{\v c}i{\'c}}, {Carilli}, {Pannella}, {Koekemoer}, {Bell}, \&
  {Salvato}}]{Karim11}
{Karim} A. {et~al.}, 2011, \apj, 730, 61

\bibitem[{{Karim} {et~al}\mbox{.}(2013){Karim}, {Swinbank}, {Hodge}, {Smail},
  {Walter}, {Biggs}, {Simpson}, {Danielson}, {Alexander}, {Bertoldi}, {de
  Breuck}, {Chapman}, {Coppin}, {Dannerbauer}, {Edge}, {Greve}, {Ivison},
  {Knudsen}, {Menten}, {Schinnerer}, {Wardlow}, {Wei{\ss}}, \& {van der
  Werf}}]{Karim12}
{Karim} A. {et~al.}, 2013, \mnras, 432, 2

\bibitem[{{Kauffmann} {et~al}\mbox{.}(2012){Kauffmann}, {Li}, {Fu},
  {Saintonge}, {Catinella}, {Tacconi}, {Kramer}, {Genzel}, {Moran}, \&
  {Schiminovich}}]{Kauff12}
{Kauffmann} G. {et~al.}, 2012, \mnras, 422, 997

\bibitem[{{Khochfar} \& {Burkert}(2006)}]{KhochfarBurkert06}
{Khochfar} S., {Burkert} A., 2006, \aap, 445, 403

\bibitem[{{Khochfar} {et~al}\mbox{.}(2011){Khochfar}, {Emsellem}, {Serra},
  {Bois}, {Alatalo}, {Bacon}, {Blitz}, {Bournaud}, {Bureau}, {Cappellari},
  {Davies}, {Davis}, {de Zeeuw}, {Duc}, {Krajnovi{\'c}}, {Kuntschner},
  {Lablanche}, {McDermid}, {Morganti}, {Naab}, {Oosterloo}, {Sarzi}, {Scott},
  {Weijmans}, \& {Young}}]{Khochfar11}
{Khochfar} S. {et~al.}, 2011, \mnras, 417, 845

\bibitem[{{Krause} {et~al}\mbox{.}(2013){Krause}, {Hirata}, {Martin}, {Neill},
  \& {Wyder}}]{Krause13}
{Krause} E., {Hirata} C.~M., {Martin} C., {Neill} J.~D., {Wyder} T.~K., 2013,
  \mnras, 428, 2548

\bibitem[{{Kravtsov}(2013)}]{Kravtsov13}
{Kravtsov} A.~V., 2013, \apjl, 764, L31

\bibitem[{{Lani} {et~al}\mbox{.}(2013){Lani}, {Almaini}, {Hartley}, {Mortlock},
  {H{\"a}u{\ss}ler}, {Chuter}, {Simpson}, {van der Wel}, {Gr{\"u}tzbauch},
  {Conselice}, {Bradshaw}, {Cooper}, {Faber}, {Grogin}, {Kocevski},
  {Koekemoer}, \& {Lai}}]{Lani13}
{Lani} C. {et~al.}, 2013, \mnras, 435, 207

\bibitem[{{Leauthaud} {et~al}\mbox{.}(2010){Leauthaud}, {Finoguenov}, {Kneib},
  {Taylor}, {Massey}, {Rhodes}, {Ilbert}, {Bundy}, {Tinker}, {George}, {Capak},
  {Koekemoer}, {Johnston}, {Zhang}, {Cappelluti}, {Ellis}, {Elvis}, {Giodini},
  {Heymans}, {Le F{\`e}vre}, {Lilly}, {McCracken}, {Mellier},
  {R{\'e}fr{\'e}gier}, {Salvato}, {Scoville}, {Smoot}, {Tanaka}, {Van
  Waerbeke}, \& {Wolk}}]{Leauthaud10}
{Leauthaud} A. {et~al.}, 2010, \apj, 709, 97

\bibitem[{{Maltby} {et~al}\mbox{.}(2010){Maltby}, {Arag{\'o}n-Salamanca},
  {Gray}, {Barden}, {H{\"a}u{\ss}ler}, {Wolf}, {Peng}, {Jahnke}, {McIntosh},
  {B{\"o}hm}, \& {van Kampen}}]{Maltby10}
{Maltby} D. {et~al.}, 2010, \mnras, 402, 282

\bibitem[{{Mandelker} {et~al}\mbox{.}(2013){Mandelker}, {Dekel}, {Ceverino},
  {Tweed}, {Moody}, \& {Primack}}]{Mandelker13}
{Mandelker} N., {Dekel} A., {Ceverino} D., {Tweed} D., {Moody} C.~E., {Primack}
  J., 2013, arXiv:1311.0013

\bibitem[{{McCavana} {et~al}\mbox{.}(2012){McCavana}, {Micic}, {Lewis},
  {Sinha}, {Sharma}, {Holley-Bockelmann}, \& {Bland-Hawthorn}}]{McCavana12}
{McCavana} T., {Micic} M., {Lewis} G.~F., {Sinha} M., {Sharma} S.,
  {Holley-Bockelmann} K., {Bland-Hawthorn} J., 2012, \mnras, 424, 361

\bibitem[{{Meert} {et~al}\mbox{.}(2013){Meert}, {Vikram}, \&
  {Bernardi}}]{Meert13}
{Meert} A., {Vikram} V., {Bernardi} M., 2013, \mnras, 433, 1344

\bibitem[{{Mei} {et~al}\mbox{.}(2012){Mei}, {Stanford}, {Holden}, {Raichoor},
  {Postman}, {Nakata}, {Finoguenov}, {Ford}, {Illingworth}, {Kodama}, {Rosati},
  {Tanaka}, {Huertas-Company}, {Rettura}, {Shankar}, {Carrasco}, {Demarco},
  {Eisenhardt}, {Jee}, {Koyama}, \& {White}}]{Mei12}
{Mei} S. {et~al.}, 2012, \apj, 754, 141

\bibitem[{{Mendel} {et~al}\mbox{.}(2013){Mendel}, {Simard}, {Ellison}, \&
  {Patton}}]{Mendel13}
{Mendel} J.~T., {Simard} L., {Ellison} S.~L., {Patton} D.~R., 2013, \mnras,
  429, 2212

\bibitem[{{Mo} {et~al}\mbox{.}(1998){Mo}, {Mao}, \& {White}}]{MMW}
{Mo} H.~J., {Mao} S., {White} S.~D.~M., 1998, \mnras, 295, 319

\bibitem[{{Mok} {et~al}\mbox{.}(2013){Mok}, {Balogh}, {McGee}, {Wilman},
  {Finoguenov}, {Tanaka}, {Giodini}, {Bower}, {Connelly}, {Hou}, {Mulchaey}, \&
  {Parker}}]{Mok13}
{Mok} A. {et~al.}, 2013, \mnras

\bibitem[{{Monaco} {et~al}\mbox{.}(2006){Monaco}, {Murante}, {Borgani}, \&
  {Fontanot}}]{Monaco06}
{Monaco} P., {Murante} G., {Borgani} S., {Fontanot} F., 2006, \apjl, 652, L89

\bibitem[{{Monaco} {et~al}\mbox{.}(2002){Monaco}, {Theuns}, \&
  {Taffoni}}]{Monaco02}
{Monaco} P., {Theuns} T., {Taffoni} G., 2002, \mnras, 331, 587

\bibitem[{{Moster} {et~al}\mbox{.}(2013){Moster}, {Naab}, \&
  {White}}]{Moster13}
{Moster} B.~P., {Naab} T., {White} S.~D.~M., 2013, \mnras, 428, 3121

\bibitem[{{Muzzin} {et~al}\mbox{.}(2012){Muzzin}, {Wilson}, {Yee}, {Gilbank},
  {Hoekstra}, {Demarco}, {Balogh}, {van Dokkum}, {Franx}, {Ellingson}, {Hicks},
  {Nantais}, {Noble}, {Lacy}, {Lidman}, {Rettura}, {Surace}, \&
  {Webb}}]{Muzzin12}
{Muzzin} A. {et~al.}, 2012, \apj, 746, 188

\bibitem[{{Naab} {et~al}\mbox{.}(2009){Naab}, {Johansson}, \&
  {Ostriker}}]{Naab09}
{Naab} T., {Johansson} P.~H., {Ostriker} J.~P., 2009, \apjl, 699, L178

\bibitem[{{Nair} {et~al}\mbox{.}(2011){Nair}, {van den Bergh}, \&
  {Abraham}}]{Nair11}
{Nair} P., {van den Bergh} S., {Abraham} R.~G., 2011, \apjl, 734, L31

\bibitem[{{Neistein} {et~al}\mbox{.}(2011){Neistein}, {Li}, {Khochfar},
  {Weinmann}, {Shankar}, \& {Boylan-Kolchin}}]{Neistein11}
{Neistein} E., {Li} C., {Khochfar} S., {Weinmann} S.~M., {Shankar} F.,
  {Boylan-Kolchin} M., 2011, \mnras, 416, 1486

\bibitem[{{Newman} {et~al}\mbox{.}(2013){Newman}, {Ellis}, {Andreon}, {Treu},
  {Raichoor}, \& {Trinchieri}}]{Newman13}
{Newman} A.~B., {Ellis} R.~S., {Andreon} S., {Treu} T., {Raichoor} A.,
  {Trinchieri} G., 2013, arXiv:1310.6754

\bibitem[{{Newman} {et~al}\mbox{.}(2012){Newman}, {Ellis}, {Bundy}, \&
  {Treu}}]{Newman12}
{Newman} A.~B., {Ellis} R.~S., {Bundy} K., {Treu} T., 2012, \apj, 746, 162

\bibitem[{{Nipoti} {et~al}\mbox{.}(2009){Nipoti}, {Treu}, {Auger}, \&
  {Bolton}}]{Nipoti09}
{Nipoti} C., {Treu} T., {Auger} M.~W., {Bolton} A.~S., 2009, \apjl, 706, L86

\bibitem[{{Nipoti} {et~al}\mbox{.}(2008){Nipoti}, {Treu}, \&
  {Bolton}}]{Nipoti08}
{Nipoti} C., {Treu} T., {Bolton} A.~S., 2008, \mnras, 390, 349

\bibitem[{{Papovich} {et~al}\mbox{.}(2012){Papovich}, {Bassett}, {Lotz}, {van
  der Wel}, {Tran}, {Finkelstein}, {Bell}, {Conselice}, {Dekel}, {Dunlop},
  {Guo}, {Faber}, {Farrah}, {Ferguson}, {Finkelstein}, {H{\"a}ussler},
  {Kocevski}, {Koekemoer}, {Koo}, {McGrath}, {McLure}, {McIntosh}, {Momcheva},
  {Newman}, {Rudnick}, {Weiner}, {Willmer}, \& {Wuyts}}]{Papovich12}
{Papovich} C. {et~al.}, 2012, \apj, 750, 93

\bibitem[{{Peeples} \& {Somerville}(2013)}]{Peeples13}
{Peeples} M.~S., {Somerville} R.~S., 2013, \mnras, 428, 1766

\bibitem[{{Poggianti} {et~al}\mbox{.}(2013){Poggianti}, {Calvi}, {Bindoni},
  {D'Onofrio}, {Moretti}, {Valentinuzzi}, {Fasano}, {Fritz}, {De Lucia},
  {Vulcani}, {Bettoni}, {Gullieuszik}, \& {Omizzolo}}]{Poggianti13}
{Poggianti} B.~M. {et~al.}, 2013, \apj, 762, 77

\bibitem[{{Posti} {et~al}\mbox{.}(2013){Posti}, {Nipoti}, {Stiavelli}, \&
  {Ciotti}}]{Posti13}
{Posti} L., {Nipoti} C., {Stiavelli} M., {Ciotti} L., 2013, arXiv:1310.2255

\bibitem[{{Prugniel} \& {Simien}(1997)}]{Prugniel97}
{Prugniel} P., {Simien} F., 1997, \aap, 321, 111

\bibitem[{{Puech} {et~al}\mbox{.}(2012){Puech}, {Hammer}, {Hopkins},
  {Athanassoula}, {Flores}, {Rodrigues}, {Wang}, \& {Yang}}]{Puech12}
{Puech} M., {Hammer} F., {Hopkins} P.~F., {Athanassoula} E., {Flores} H.,
  {Rodrigues} M., {Wang} J.~L., {Yang} Y.~B., 2012, \apj, 753, 128

\bibitem[{{Raichoor} {et~al}\mbox{.}(2012){Raichoor}, {Mei}, {Stanford},
  {Holden}, {Nakata}, {Rosati}, {Shankar}, {Tanaka}, {Ford}, {Huertas-Company},
  {Illingworth}, {Kodama}, {Postman}, {Rettura}, {Blakeslee}, {Demarco}, {Jee},
  \& {White}}]{Raichoor12}
{Raichoor} A. {et~al.}, 2012, \apj, 745, 130

\bibitem[{{Rettura} {et~al}\mbox{.}(2010){Rettura}, {Rosati}, {Nonino},
  {Fosbury}, {Gobat}, {Menci}, {Strazzullo}, {Mei}, {Demarco}, \&
  {Ford}}]{Rettura10}
{Rettura} A. {et~al.}, 2010, \apj, 709, 512

\bibitem[{{Rodr{\'{\i}}guez-Puebla}
  {et~al}\mbox{.}(2012){Rodr{\'{\i}}guez-Puebla}, {Drory}, \&
  {Avila-Reese}}]{Rodriguez12}
{Rodr{\'{\i}}guez-Puebla} A., {Drory} N., {Avila-Reese} V., 2012, \apj, 756, 2

\bibitem[{{Saglia} {et~al}\mbox{.}(1997){Saglia}, {Bertschinger}, {Baggley},
  {Burstein}, {Colless}, {Davies}, {McMahan}, \& {Wegner}}]{Saglia97}
{Saglia} R.~P., {Bertschinger} E., {Baggley} G., {Burstein} D., {Colless} M.,
  {Davies} R.~L., {McMahan}, Jr. R.~K., {Wegner} G., 1997, \apjs, 109, 79

\bibitem[{{S{\'e}rsic}(1963)}]{Sersic63}
{S{\'e}rsic} J.~L., 1963, Boletin de la Asociacion Argentina de Astronomia La
  Plata Argentina, 6, 41

\bibitem[{{Shankar} \& {Bernardi}(2009)}]{ShankarBernardi09}
{Shankar} F., {Bernardi} M., 2009, \mnras, 396, L76

\bibitem[{{Shankar} {et~al}\mbox{.}(2006){Shankar}, {Lapi}, {Salucci}, {De
  Zotti}, \& {Danese}}]{Shankar06}
{Shankar} F., {Lapi} A., {Salucci} P., {De Zotti} G., {Danese} L., 2006, \apj,
  643, 14

\bibitem[{{Shankar} {et~al}\mbox{.}(2010{\natexlab{a}}){Shankar}, {Marulli},
  {Bernardi}, {Boylan-Kolchin}, {Dai}, \& {Khochfar}}]{ShankarPhire}
{Shankar} F., {Marulli} F., {Bernardi} M., {Boylan-Kolchin} M., {Dai} X.,
  {Khochfar} S., 2010{\natexlab{a}}, \mnras, 405, 948

\bibitem[{{Shankar} {et~al}\mbox{.}(2010{\natexlab{b}}){Shankar}, {Marulli},
  {Bernardi}, {Dai}, {Hyde}, \& {Sheth}}]{ShankarRe}
{Shankar} F., {Marulli} F., {Bernardi} M., {Dai} X., {Hyde} J.~B., {Sheth}
  R.~K., 2010{\natexlab{b}}, \mnras, 403, 117

\bibitem[{{Shankar} {et~al}\mbox{.}(2013){Shankar}, {Marulli}, {Bernardi},
  {Mei}, {Meert}, \& {Vikram}}]{Shankar13}
{Shankar} F., {Marulli} F., {Bernardi} M., {Mei} S., {Meert} A., {Vikram} V.,
  2013, \mnras, 428, 109

\bibitem[{{Shen} {et~al}\mbox{.}(2003){Shen}, {Mo}, {White}, {Blanton},
  {Kauffmann}, {Voges}, {Brinkmann}, \& {Csabai}}]{Shen03}
{Shen} S., {Mo} H.~J., {White} S.~D.~M., {Blanton} M.~R., {Kauffmann} G.,
  {Voges} W., {Brinkmann} J., {Csabai} I., 2003, \mnras, 343, 978

\bibitem[{{Somerville} {et~al}\mbox{.}(2008){Somerville}, {Barden}, {Rix},
  {Bell}, {Beckwith}, {Borch}, {Caldwell}, {H{\"a}u{\ss}ler}, {Heymans},
  {Jahnke}, {Jogee}, {McIntosh}, {Meisenheimer}, {Peng}, {S{\'a}nchez},
  {Wisotzki}, \& {Wolf}}]{Somerville08}
{Somerville} R.~S. {et~al.}, 2008, \apj, 672, 776

\bibitem[{{Somerville} {et~al}\mbox{.}(2001){Somerville}, {Primack}, \&
  {Faber}}]{Somerville01}
{Somerville} R.~S., {Primack} J.~R., {Faber} S.~M., 2001, \mnras, 320, 504

\bibitem[{{Springel}(2005)}]{Springel05}
{Springel} V., 2005, \mnras, 364, 1105

\bibitem[{{Stewart} {et~al}\mbox{.}(2009){Stewart}, {Bullock}, {Wechsler}, \&
  {Maller}}]{Stewart09}
{Stewart} K.~R., {Bullock} J.~S., {Wechsler} R.~H., {Maller} A.~H., 2009, \apj,
  702, 307

\bibitem[{{Strazzullo} {et~al}\mbox{.}(2013){Strazzullo}, {Gobat}, {Daddi},
  {Onodera}, {Carollo}, {Dickinson}, {Renzini}, {Arimoto}, {Cimatti},
  {Finoguenov}, \& {Chary}}]{Strazzu13}
{Strazzullo} V. {et~al.}, 2013, \apj, 772, 118

\bibitem[{{Stringer} {et~al}\mbox{.}(2013){Stringer}, {Shankar}, {Novak},
  {Huertas-Company}, {Combes}, \& {Moster}}]{Stringer13}
{Stringer} M.~J., {Shankar} F., {Novak} G.~S., {Huertas-Company} M., {Combes}
  F., {Moster} B.~P., 2013, arXiv:1310.3823

\bibitem[{{Taffoni} {et~al}\mbox{.}(2003){Taffoni}, {Mayer}, {Colpi}, \&
  {Governato}}]{Taffoni03}
{Taffoni} G., {Mayer} L., {Colpi} M., {Governato} F., 2003, \mnras, 341, 434

\bibitem[{{Taranu} {et~al}\mbox{.}(2013){Taranu}, {Dubinski}, \&
  {Yee}}]{Taranu13}
{Taranu} D.~S., {Dubinski} J., {Yee} H.~K.~C., 2013, \apj, 778, 61

\bibitem[{{Vale} \& {Ostriker}(2004)}]{Vale04}
{Vale} A., {Ostriker} J.~P., 2004, \mnras, 353, 189

\bibitem[{{Valentinuzzi} {et~al}\mbox{.}(2010{\natexlab{a}}){Valentinuzzi},
  {Fritz}, {Poggianti}, {Cava}, {Bettoni}, {Fasano}, {D'Onofrio}, {Couch},
  {Dressler}, {Moles}, {Moretti}, {Omizzolo}, {Kj{\ae}rgaard}, {Vanzella}, \&
  {Varela}}]{Valentinuzzi10a}
{Valentinuzzi} T. {et~al.}, 2010{\natexlab{a}}, \apj, 712, 226

\bibitem[{{Valentinuzzi} {et~al}\mbox{.}(2010{\natexlab{b}}){Valentinuzzi},
  {Poggianti}, {Saglia}, {Arag{\'o}n-Salamanca}, {Simard},
  {S{\'a}nchez-Bl{\'a}zquez}, {D'onofrio}, {Cava}, {Couch}, {Fritz}, {Moretti},
  \& {Vulcani}}]{Valentinuzzi10b}
{Valentinuzzi} T. {et~al.}, 2010{\natexlab{b}}, \apjl, 721, L19

\bibitem[{{van der Wel} {et~al}\mbox{.}(2009){van der Wel}, {Bell}, {van den
  Bosch}, {Gallazzi}, \& {Rix}}]{vanderwel09}
{van der Wel} A., {Bell} E.~F., {van den Bosch} F.~C., {Gallazzi} A., {Rix} H.,
  2009, \apj, 698, 1232

\bibitem[{{van Dokkum} {et~al}\mbox{.}(2010){van Dokkum}, {Whitaker},
  {Brammer}, {Franx}, {Kriek}, {Labb{\'e}}, {Marchesini}, {Quadri}, {Bezanson},
  {Illingworth}, {Muzzin}, {Rudnick}, {Tal}, \& {Wake}}]{VanDokkum10}
{van Dokkum} P.~G. {et~al.}, 2010, \apj, 709, 1018

\bibitem[{{Vikram} {et~al}\mbox{.}(2010){Vikram}, {Wadadekar}, {Kembhavi}, \&
  {Vijayagovindan}}]{Vikram10}
{Vikram} V., {Wadadekar} Y., {Kembhavi} A.~K., {Vijayagovindan} G.~V., 2010,
  \mnras, 409, 1379

\bibitem[{{Watson} {et~al}\mbox{.}(2012){Watson}, {Berlind}, \&
  {Zentner}}]{Watson12}
{Watson} D.~F., {Berlind} A.~A., {Zentner} A.~R., 2012, \apj, 754, 90

\bibitem[{{Weinmann} {et~al}\mbox{.}(2009){Weinmann}, {Kauffmann}, {van den
  Bosch}, {Pasquali}, {McIntosh}, {Mo}, {Yang}, \& {Guo}}]{Weinmann09}
{Weinmann} S.~M., {Kauffmann} G., {van den Bosch} F.~C., {Pasquali} A.,
  {McIntosh} D.~H., {Mo} H., {Yang} X., {Guo} Y., 2009, \mnras, 394, 1213

\bibitem[{{Wetzel} {et~al}\mbox{.}(2013){Wetzel}, {Tinker}, {Conroy}, \& {van
  den Bosch}}]{Wetzel13}
{Wetzel} A.~R., {Tinker} J.~L., {Conroy} C., {van den Bosch} F.~C., 2013,
  \mnras, 432, 336

\bibitem[{{Wilman} {et~al}\mbox{.}(2013){Wilman}, {Fontanot}, {De Lucia},
  {Erwin}, \& {Monaco}}]{Wilman13}
{Wilman} D.~J., {Fontanot} F., {De Lucia} G., {Erwin} P., {Monaco} P., 2013,
  \mnras

\bibitem[{{Woo} {et~al}\mbox{.}(2013){Woo}, {Dekel}, {Faber}, {Noeske}, {Koo},
  {Gerke}, {Cooper}, {Salim}, {Dutton}, {Newman}, {Weiner}, {Bundy}, {Willmer},
  {Davis}, \& {Yan}}]{Woo13}
{Woo} J. {et~al.}, 2013, \mnras, 428, 3306

\bibitem[{{Yang} {et~al}\mbox{.}(2013){Yang}, {Mo}, {van den Bosch}, {Bonaca},
  {Li}, {Lu}, {Lu}, \& {Lu}}]{Yang13}
{Yang} X., {Mo} H.~J., {van den Bosch} F.~C., {Bonaca} A., {Li} S., {Lu} Y.,
  {Lu} Y., {Lu} Z., 2013, \apj, 770, 115

\bibitem[{{Yang} {et~al}\mbox{.}(2007){Yang}, {Mo}, {van den Bosch},
  {Pasquali}, {Li}, \& {Barden}}]{Yang07}
{Yang} X., {Mo} H.~J., {van den Bosch} F.~C., {Pasquali} A., {Li} C., {Barden}
  M., 2007, \apj, 671, 153

\bibitem[{{Yang} {et~al}\mbox{.}(2012){Yang}, {Mo}, {van den Bosch}, {Zhang},
  \& {Han}}]{Yang12}
{Yang} X., {Mo} H.~J., {van den Bosch} F.~C., {Zhang} Y., {Han} J., 2012, \apj,
  752, 41

\bibitem[{{Zavala} {et~al}\mbox{.}(2012){Zavala}, {Avila-Reese}, {Firmani}, \&
  {Boylan-Kolchin}}]{Zavala12}
{Zavala} J., {Avila-Reese} V., {Firmani} C., {Boylan-Kolchin} M., 2012, \mnras,
  427, 1503

\end{thebibliography}

\appendix

\section{Gas Fraction in the Local Universe}
\label{app|gasfractions}

\begin{figure}
    \includegraphics[width=8.5truecm]{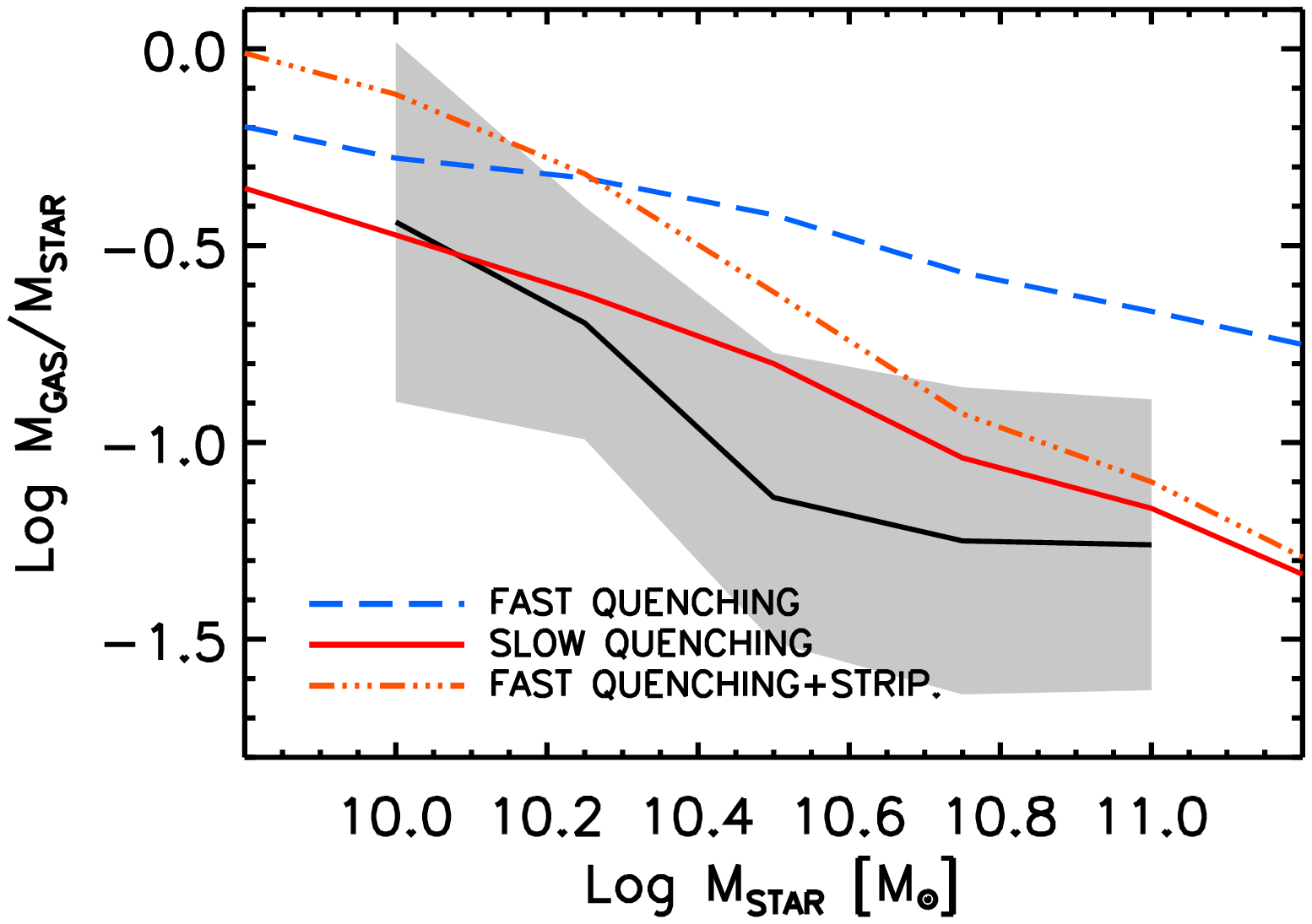}
    \caption{Predicted median gas fraction of bulge dominated galaxies
    for the models with continuous star formation in satellites (\emph{solid}, \emph{red}) and minimal satellites growth after infall (\emph{long-dashed}, \emph{red}),
    compared with the data by \citealt{Catinella10} from the GASS survey
    for early-type, bulge dominated galaxies (\emph{solid} line with \emph{grey area} marking the 1-$\sigma$ in the full distribution).}
    \label{fig|Gasfractions}
\end{figure}

We discussed in \sect\ref{subsec|PhysicsOfEnvironment} that models with stronger
stripping tend to predict smaller sizes and more pronounced environmental dependence than those without stripping.
We here add that this type of models also produce somewhat
different outcomes for the gas fractions in the remnant massive ellipticals.
This is clearly evident in \figu\ref{fig|Gasfractions},
where we compare the resulting gas fractions as a function of stellar mass
against the data from the GASS survey by \citealt{Catinella10} for early-type galaxies
(solid line with grey area).
The model with fast quenching would produce too large gas fractions retained in the
remnant galaxies (blue/long-dashed line) compared to the slow quenching one (solid/red line),
in which a significant part of the gas in the merging satellites is consumed during infall.
Clearly efficient gas stripping must accompany the fast quenching model (orange/triple dot-dashed line)
to reconcile model predictions with observations. The latter model is characterized by the same value of
$\eta=0.25$ (\eq\ref{eq|stripping}), for both the stars and the gas component, as the stripping model reported in \figu\ref{fig|MstarMhaloRel}.
Overall, a fast quenching+stripping model is quite degenerate with a slow quenching model.
However, the former tends to produce too compact remnants at fixed stellar mass and stronger environmental
dependence with respect to observations.

\section{Satellites}
\label{App|Satellites}

So far we focused our attention on central galaxies, as these are the systems
for which the effect of mergers is expected to be maximized, being the satellite-satellite merger
rate measured to be very low in dark matter simulations \citep[e.g.,][]{Angulo09}.
For completeness, this Appendix is dedicated
to briefly explore and compare the main observed and predicted structural properties of satellites.

\begin{figure*}
    \includegraphics[width=15truecm]{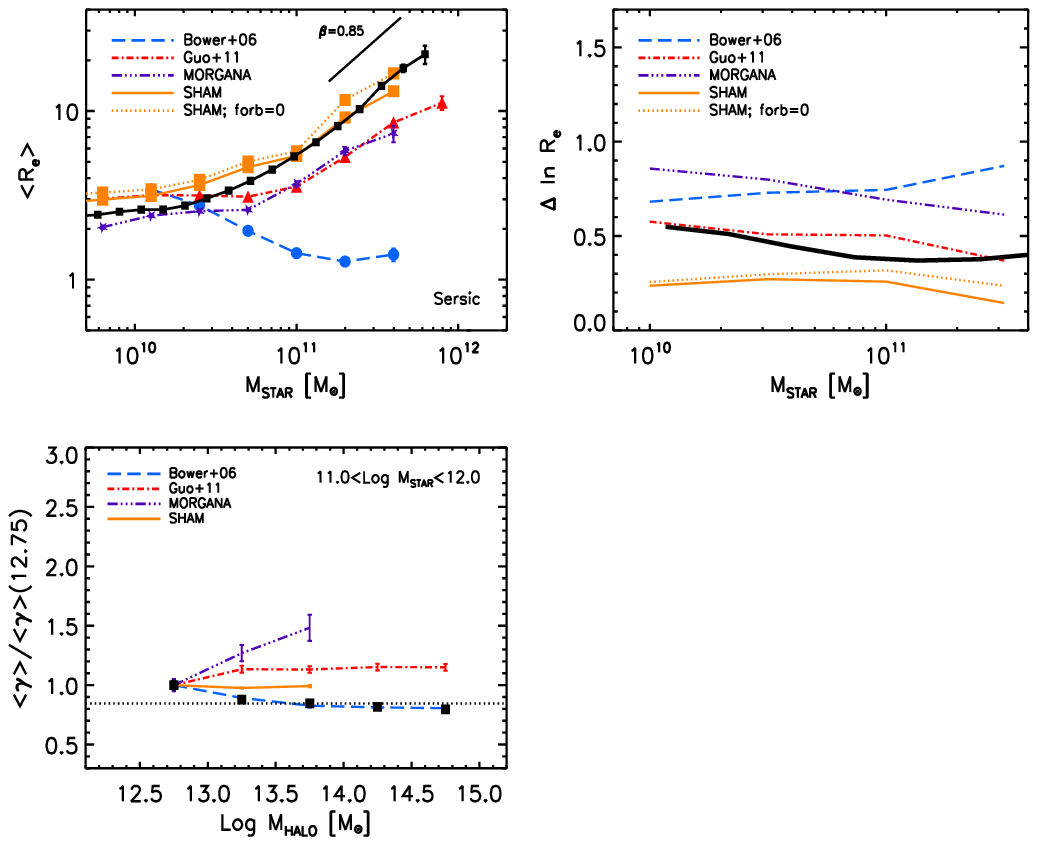}
    \caption{Predicted structural properties of satellites. \emph{Top left}: size-stellar mass relation
    in different models, as labelled, compared to data from \citep[][\emph{solid}, \emph{black} line]{Huertas13b}.
    \emph{Top right}: predicted scatter around the size-stellar mass relation for the same models and data as in the left panel.
    \emph{Bottom left}: median normalized sizes \gammar, divided by the value at $\log \mhaloe=12.5$; note that
    only galaxies with $B/T>0.5$ have been selected here, to make better contact with the data and previous discussion
    on central galaxies, but similar results are found even when no cut in $B/T$ is imposed.}
    \label{fig|Satellites}
\end{figure*}

\figu\ref{fig|Satellites} is a collection of three panels. The top, left panel reports the
expected size-stellar mass relation from different models, as labelled, for all satellite
galaxies with no restriction in the bulge component. We here apply the 3D-to-2D correction
only for bulge-dominated galaxies with \bt$>0.5$, while leave the sizes of disc-dominated
galaxies unaltered \citep[e.g.,][]{Kravtsov13}. Both the data and the models show a clear flattening
in median size at low masses, below $\mstare \lesssim 10^{11}\, \msune$. As evidenced from the two component
fitting of SDSS galaxies by \citet{Bernardi13}, the latter feature is naturally explained
by the growing contribution of discs progressively dominating the structural properties of galaxies
at lower stellar masses. What is most relevant here is that, except for the B06 model,
all models share size distributions for satellites in broad agreement with those observed. In other words, the
satellites, which eventually will merge with their centrals, have the correct sizes.

The upper right panel plots the scatter around the median relation for the same models compared with the data (solid line).
As for centrals, models tend to predict larger distributions, although the difference is limited
for the G11 model and absent for the SHAM model, both characterized by low disc instabilities and, at least the latter,
by tight correlations for the infalling satellites (see \sect\ref{subsec|ScatterReMstarAndBTfractions}).

The lower left panel shows that the environmental dependence of satellites is overall quite limited by up to $20-30\%$.
Here only raw model predictions are reported, the convolution with (correlated) errors will clearly cancel these relatively
modest trends with environment. Overall, satellites tend to present somewhat less environmental dependence with respect
to centrals of similar stellar mass (\figu\ref{fig|ReMstarFixedMhaloNorm}).
Particularly striking is the difference from centrals to satellites in the B06 model, for which the dependence is actually reversed.
Satellites are a mixture of galaxies with different morphologies and accretion histories.
Their varied evolutionary histories bring them to live in very disparate environments, much more than their counterpart centrals of
similar stellar mass, thus decreasing any environmental signal \citep[see also][]{Huertas13b}.

\label{lastpage}
\end{document}